\documentclass[journal]{IEEEtran}

\ifCLASSINFOpdf

\else

\fi

\usepackage{array}
\usepackage{amsthm,amsmath,amssymb}
\usepackage{mathrsfs}
\usepackage{multirow}
\usepackage{siunitx}

\usepackage{booktabs}
\usepackage{graphicx}
 \usepackage{enumerate}
\usepackage{cite}
\usepackage{tabu}
\usepackage{color}
\usepackage{subfigure}
\begin{document}
\newcommand{\ee}{\end{equation}}
\newcommand{\br}{{\mbox{\boldmath{$r$}}}}
\newcommand{\bp}{{\mbox{\boldmath{$p$}}}}
\newcommand{\bpi}{\mbox{\boldmath{ $\pi $}}}
\newcommand{\bn}{{\mbox{\boldmath{$n$}}}}
\newcommand{\balfa}{{\mbox{\boldmath{$\alpha$}}}}
\newcommand{\ba}{\mbox{\boldmath{$a $}}}
\newcommand{\bta}{\mbox{\boldmath{$\beta $}}}
\newcommand{\bg}{\mbox{\boldmath{$g $}}}
\newcommand{\bPsi}{\mbox{\boldmath{$\Psi $}}}
\newcommand{\bsigma}{\mbox{\boldmath{ $\Sigma $}}}
\newcommand{\bGamma}{{\bf \Gamma }}
\newcommand{\bA}{{\bf A }}
\newcommand{\bP}{{\bf P }}
\newcommand{\bX}{{\bf X }}
\newcommand{\bI}{{\bf I }}
\newcommand{\bR}{{\bf R }}
\newcommand{\bZ}{{\bf Z }}
\newcommand{\bz}{{\bf z }}
\newcommand{\bx}{{\mathbf{x}}}
\newcommand{\bM}{{\bf M}}
\newcommand{\bU}{{\bf U}}
\newcommand{\bD}{{\bf D}}
\newcommand{\bJ}{{\bf J}}
\newcommand{\bH}{{\bf H}}
\newcommand{\bK}{{\bf K}}
\newcommand{\bm}{{\bf m}}
\newcommand{\bN}{{\bf N}}
\newcommand{\bC}{{\bf C}}
\newcommand{\bL}{{\bf L}}
\newcommand{\bF}{{\bf F}}
\newcommand{\bv}{{\bf v}}
\newcommand{\bSigma}{{\bf \Sigma}}
\newcommand{\bS}{{\bf S}}
\newcommand{\bs}{{\bf s}}
\newcommand{\bO}{{\bf O}}
\newcommand{\bQ}{{\bf Q}}
\newcommand{\btr}{{\mbox{\boldmath{$tr$}}}}
\newcommand{\bNSCM}{{\bf NSCM}}
\newcommand{\barg}{{\bf arg}}
\newcommand{\bmax}{{\bf max}}
\newcommand{\test}{\mbox{$
	\begin{array}{c}
	\stackrel{ \stackrel{\textstyle H_1}{\textstyle >} } { \stackrel{\textstyle <}{\textstyle H_0} }
	\end{array}
	$}}
\newcommand{\tabincell}[2]{\begin{tabular}{@{}#1@{}}#2\end{tabular}}
\newtheorem{Def}{Definition}
\newtheorem{Pro}{Proposition}
\newtheorem{Lem}{Lemma}
\newtheorem{Exa}{Example}
\newtheorem{Rem}{Remark}
\newtheorem{Cor}{Corollary}
\renewcommand{\labelitemi}{$\bullet$}

\title{Trajectory PHD and CPHD Filters with \\ Unknown Detection Profile}

\author{Shaoxiu~Wei,
        Boxiang~Zhang,
        and~Wei~Yi
        \thanks{S. Wei, B. Zhang, and W. Yi are with the School of Information and Communication Engineering, University of Electronic Science and Technology of China.
}}

\maketitle

\begin{abstract}
Compared to the probability hypothesis density (PHD) and cardinalized PHD (CPHD) filters, the trajectory PHD (TPHD) and {\color{black}trajectory CPHD (TCPHD)} filters are for sets of trajectories, and thus are able to produce trajectory estimates with better estimation performance.
In this paper, we develop the TPHD and TCPHD filters which can adaptively learn the history of the unknown target detection probability, and therefore they can perform more robustly in scenarios where targets are with unknown and time-varying detection probabilities. These filters are referred to as the unknown TPHD (U-TPHD) and unknown TCPHD (U-TCPHD) filters.
By minimizing the Kullback-Leibler divergence (KLD), the U-TPHD and U-TCPHD filters can obtain, respectively, the best Poisson and independent identically distributed (IID) density approximations over the augmented sets of trajectories. For computational efficiency, we also propose the U-TPHD and U-TCPHD filters that only consider the unknown detection profile at the current time. Specifically, the Beta-Gaussian mixture method is adopted for the implementation of proposed filters, which are referred to as the BG-U-TPHD and BG-U-TCPHD filters. The $L$-scan approximations of these filters with much lower computational burden are also presented. Finally, various simulation results demonstrate that the BG-U-TPHD and BG-U-TCPHD filters can achieve robust tracking performance to adapt to unknown detection profile. Besides, it also shows that usually a small value of the $L$-scan approximation can achieve almost full efficiency of both filters but with a much lower computational costs.
\end{abstract}

\IEEEpeerreviewmaketitle

\section{Introduction}
\IEEEPARstart{T}{he} purpose of multi-target tracking is to estimate the time-varying number and states of targets through a set of measurements in the presence of data association uncertainty, detection uncertainty, false measurements, and noise \cite{blackman1986book,MTT-TVT-Dai,MTT-TVT-Savic,MTT-TVT-WYi,MTT-TVT-Yeow}. There are three main approaches to multi-target tracking: the joint probabilistic data association (JPDA) filter \cite{JPDA-1998,JPDA-1983}, the multiple hypotheses tracking (MHT) \cite{blackman1986book,blackman2004MHT} and the random finite set (RFS)\cite{Mahler2007RFSbook}.
Among them, the RFS approaches aim to model the appearance and disappearance of targets, misdetections and false alarms within a unified Bayesian framework~\cite{Mahler2007RFSbook}.

Several tractable and useful multi-object filters have been developed based on RFS methods, including the probability hypothesis density (PHD) filter~\cite{Mahler2007RFSbook,Mahler2003RFS,Vo2006PHD}, the cardinalized PHD (CPHD) filter \cite{Mahler2007RFSbook,Mahler2007CPHD,Vo2007CPHD}, the multi-Bernoulli (MB) filter \cite{Mahler2007RFSbook,Vo2009MB}, the Poisson multi-Bernoulli mixture (PMBM) filter \cite{Williams2015PMBM,Angel2018PMBM}, the generalized labeled multi-Bernoulli (GLMB) filter \cite{Vo2013GLMB,Vo2014GLMB}, and the labeled multi-Bernoulli (LMB) filter \cite{Reuter2014LMB}. 
Among them, the PHD and CPHD filters are the most fundamental RFS filters and known for their low computational burden. The PHD filter considers a Poisson multi-target filtering density, while the CPHD filter considers an independent and identically distributed (IID) cluster multi-target filtering density. If the prior or posterior density is not Poisson or IID cluster, the PHD and CPHD filters can be achieved using the best Poisson or IID cluster approximations that minimizes the Kullback-Leibler divergence (KLD) \cite{Angel2015KLD}. {\color{black}While usually the Poisson prior in the PHD filter can be directly achieved by the Poisson input when there is no spawning and the RFS of new born targets is also Poisson.} Both filters can be implemented efficiently using numerical solutions such as sequential Monte Carlo (SMC) \cite{Vo2005smc,Ristic2010smc}, or Gaussian mixture (GM)~\cite{Vo2006PHD,Vo2007CPHD}. In recent years, the PHD and CPHD filters have also been successfully applied to the distributed multi-sensor fusion \cite{Battistelli2013Fusion,Uney2013Fusion}, robotics \cite{Gao2013Robotic,Mullane2011Robitc} and visual tracking \cite{Maggio2008visual,Zhou2014visual}.

All the aforementioned filters use a set of targets as the state variable, and update the filtering density at each measurement time.
Since the filtering density only captures the statistical information on the current multi-target state variable, the history estimates cannot be updated with the subsequent measurements. As a result, these filtering density based filters usually have poorer performance compared to the smoothing methods~\cite{Vo2019MsGLMB}. Besides, without the assistant of the target labels (or tags), the unlabeled filters such as the PHD and CPHD filters can not form target trajectories.

Recently, the trajectory PHD (TPHD) and trajectory cardinalized PHD (TCPHD) filters are proposed by adopting a set of trajectories as the state variable~\cite{Svensson2014TMTT,Angel2019TPHD}. In this way, the TPHD and TCPHD filters are able to establish trajectories directly and deliver better estimation performance than the standard PHD and CPHD filters. Specifically, based on the KLD minimization, the TPHD and TCPHD filters propagate the best approximated PHD of the Poisson and IID cluster multi-trajectory densities respectively. The TPHD and TCPHD filters do not marginalize the past target states in the prediction step of PHD and are able to update the whole trajectory in the update step using the current measurements. This is also the reason why the TPHD and TCPHD filters can outperform the standard PHD and CPHD filters. In~\cite{Angel2019TPHD}, the Gaussian mixture (GM) implementations for the TPHD and TCPHD filters are also proposed and referred to as GM-TPHD and GM-TCPHD. Meanwhile, the $L$-scan approximation strategy is suggested to achieve the fast implementation by only updating the multi-trajectory density of the last $L$ time. Following the same routine, the trajectory MB (TMB) filter and the trajectory PMBM (TPMBM) filter {\color{black}have also been devised} recently~\cite{Angel2020TMB,Granstrom2018TPMBM}.
In~\cite{Vo2019MsGLMB}, by adopting a set of trajectories as the state variable, a multi-scan version of the GLMB filter is proposed to directly propagates the labeled multi-trajectory posterior density. The multi-scan GLMB recursion does not marginalize over the past label sets and the past trajectory states, and in principle the multi-trajectory posterior contains the complete statistical characterization of the interested variables based on all the historical measurement sets. Consequently, the multi-scan GLMB filter can provide excellent estimation performance of target trajectories. On the other hand, techniques such as component truncation and Gibbs sampler need to be employed to enable an efficient implementation~\cite{Vo2013GLMB,Vo2014GLMB,Vo2019MsGLMB}.

In the multi-target tracking, the knowledge of two uncertain sources, namely, the clutter rate and detection probability are also of significant importance. However, in many practical applications, {\color{black} the clutter rate and detection probability change dynamically between time steps in unpredictable ways.} In general, it is difficult to infer these characteristics from training data. Rather they should be inferred online from the actual measurement-stream~\cite{Mahler2009UB}. To deal with the issues of unknown clutter rate, the robust PHD/CPHD filters based on a clutter generator have been proposed~\cite{Chen2012UC,Mahler2009UB,Mahler2014UB,Mahler2011UPHD}. These methods assume that the multi-target state consists of both the finite set of actual targets and clutter generators, and can obtain clutter information by tracking the clutter generators~\cite{Mahler2011UPHD}. As for the unknown detection probability, several improved filters have been proposed to adaptively estimate the unknown detection profile \cite{Papa2016Robust,GLi2020RPMBM,Punchihewa2018RGLMB,Vo2013RMB,Mahler2011UPHD,CLi2018UPHD}.
For example, the augmented state model is proposed to adopt the unknown detection profile using the Beta-Gaussian mixtures in the PHD and CPHD filters~\cite{Mahler2011UPHD}. Later, the inverse Gamma-Gaussian mixture is employed to propagate
non-negative features, including signal amplitude and SNR,
which are non-Gaussian~\cite{CLi2018UPHD}.

In this paper, we aim to develop the TPHD and TCPHD filters which can perform robustly in scenarios with unknown and time-varying target detection probability, and they are referred to as the unknown TPHD (U-TPHD) and unknown TCPHD (U-TCPHD) filters. Specifically, during the derivations of U-TPHD and U-TCPHD, {\color{black}the Beta-Gaussian mixture approach~\cite{Mahler2011UPHD} is chosen} to model the detection probability. The robust version of both filters can be devised similarly using the inverse Gamma-Gaussian mixtures~\cite{CLi2018UPHD} but are not considered here. The main contributions of this paper are given as follows:
\begin{enumerate}
	\item \emph{Derivations of the recursive equations for the U-TPHD/U-TCPHD filters}: Generally, the TPHD and TCPHD filters propagate forward the PHD of Poisson and IID cluster multi-trajectory densities, respectively. In principle, by augmenting the sequence of the unknown detection probabilities into the set of trajectories, the filters can not only adaptively learn the uneven detection profile, but also obtain the augmented target trajectories. However, the predicted or the updated densities need not to be Poisson or IID cluster multi-trajectory densities. Therefore, we propose to obtain the best Poisson/IID cluster approximation of the multi-trajectory densities over the augmented sets of trajectories by minimizing the KLD. Besides, considering computation efficiency, the analytic recursions are also presented of both the U-TPHD and U-TCPHD filters, which consider the effect of the unknown detection probability for only the latest frame time. 
	\item \emph{The Beta-Gaussian mixture implementations for the U-TPHD and U-TCPHD filters}: In the algorithmic implementation, the unknown detection probability is modeled by using the Beta-Gaussian (BG) mixtures, where each component is a product of a Beta density (on the augmented part) and a Gaussian density (on the trajectory)~\cite{Mahler2011UPHD}. The resulting filters are named as the BG-U-TPHD and BG-U-TCPHD filters for short. In consideration of the algorithmic complexity, we propose the approximations of the BG-U-TPHD and BG-U-TCPHD filters that only consider the unknown detection profile at current time. The $L$-scan approximations of these filters are proposed to achieve lower computational costs. Simulation results demonstrate that both filters can achieve excellent performance in the scenarios where the targets are with unknown and time-varying detection probabilities.
\end{enumerate}

The remainder of the paper is organized as follows. Section II presents the background materials on the TPHD and TCPHD filters. In Section III, the detail derivations of the proposed U-TPHD and U-TCPHD filters are given. Their Beta-Gaussian mixture implementations and the corresponding $L$-scan approximations are developed in Section IV. The estimation step including pruning and absorption procedures is also delineated in this section. The performance assessment of the proposed filters are given in Section V. Lastly, conclusions are drawn in Section VI.

\section{Background}
This section provides a brief review of the trajectory RFS, the TPHD and TCPHD filters. Notations are first given in Section II-A. In Section II-B, the Bayesian recursions for both the TPHD and TCPHD filters are reviewed. The prediction and update steps of the TPHD and TCPHD filters are given in Section II-C. Further details are available in~\cite{Angel2019TPHD}.
\subsection{Notations}
The variable~$X=(t,x^{1:i})$ with $x\in \mathbb{R}^{n_x}$ is used to denote a single trajectory, where $t$ represents its birth time, $n_x$ is the dimension of target state $x$, and $x^{1:i}=(x^1,...,x^i)$ denotes a sequence including the target states at each {\color{black}time step} of the trajectory with length $i$. At time $k$, the trajectory state space is defined as
\begin{align}
	\mathbb T_k=\uplus_{(t,i)\in \mathbb{I}_k}\{t\}\times\mathbb{R}^{in_x},
\label{eq:T_space}
\end{align}
where $\uplus$ denotes the disjoint union, $\times$ denotes a Cartesian product, and $\mathbb{I}_k$ is a discrete variable state space that equals to $\{(t,i):1\leq t\leq k~\text{and}~1\leq i\leq k-t+1\}$. A single trajectory density is given as $\bar{p}(X)$ and its integral is expressed as \cite{Angel2020TMTT}
\begin{align}
	\int_\mathbb{T}{\bar p\left(X\right)dX}=\sum_{(t,i)\in \mathbb{I}}{\int_{\mathbb{R}^{in_x}}\bar p\left(t,x^{1:i}\right)dx^{1:i}}.
\end{align}
Similar to the target RFS, the trajectory RFS at time $k$ is defined as
\begin{align}
\mathbf{X}_k=\{X_{1},...,X_{N^k}\}\in \mathcal{F}(\mathbb T_k),
\end{align}
where $\mathcal{F}(\mathbb T_k)$ is the respective collections of all finite subsets of $\mathbb T_k$ defined by \eqref{eq:T_space}. The PHD of posterior multi-trajectory density at time $k$ is denoted as $D_k(X)$. Given a trajectory $X$, the corresponding target state at time $k$ is given as $\tau_k(X)$. Similarly, given the multi-trajectory state $\mathbf{X}_k$, the multi-target state is $\tau_k(\mathbf{X}_k)=\cup_{X\in\mathbf{X}_k}\tau_k(X)$.

The single-trajectory state is denoted using the uppercase letter (e.g. $X$ ), whose subscript only represents the number (e.g. $X_j$). The multi-trajectory state is represented by the bold letter (e.g. $\mathbf X$), whose subscript only represents the time (e.g. $\mathbf {X}_k$). The binomial coefficient and the permutation coefficient notations are given as, respectively,
$
C_j^\ell  = \frac{{\ell !}}{{j!\left( {\ell  - j} \right)!}}$ and
$P_j^\ell  = \frac{{\ell !}}{{\left( {\ell  - j} \right)!}}.
$
The generalized Kronecker delta function is given by
\begin{align}
		\delta_{A}(B) \triangleq \left\{
	\begin{array}{lr}
		1, ~~\text{if}~A=B& \\
		0, ~~\text{otherwise}.&
	\end{array}
	\right.
\end{align}
for continuous variables {\color{black}and $\delta_{A}[B]$ represents for discrete variables.} Besides, the inner product between two real valued functions $a$ and $b$ is expressed as $\langle {a,b}\rangle$, which equals to $\int {a( x )b(x)} dx$, and we have $\langle {a,b}\rangle  = \begin{matrix} \sum_{n=0}^\infty \end{matrix}  {a( n )b( n)}$, when $a$ and $b$ are both real sequences. For a finite set $\mathcal{Z}$ of real numbers, its \emph{elementary symmetric function} \cite{Borwein1995book} of order $q$ is given as
\begin{align}
{e_q}\left( \mathcal{Z} \right) = \sum\limits_{\sigma \subseteq \mathcal{Z},\left| \sigma \right| = q} {\prod\limits_{\xi  \in \sigma} \xi  },
\end{align}
and the subtraction of sets is represented by the notation $\backslash$.
\subsection{Bayesian Filtering Recursion}
The multi-trajectory state and set of measurements at the time step $k$ are the finite sets expressed as follows,
\begin{align}
	\mathbf{X}_k=&\{X_{1},...,X_{N^k}\}\in \mathcal{F}(\mathbb T_k),\\
	{Z}_k=&\{z_{1},...,z_{M^k}\}\in \mathcal{F}(\mathbb{Z}_k).
\end{align}
where the $N^k$ trajectories {\color{black}take} values in the state space $\mathbb T_k$, the $M^k$ measurements {\color{black}take} values in the measurement space $\mathbb{Z}_k\triangleq \mathbb{R}^{n_z}$.

If the posterior multi-trajectory density $p_{k-1}$ at time $k-1$ is given, the posterior density at time $k$ can be computed by using the Bayes recursion \cite{Angel2019TPHD}
\begin{align}
\ p_{k|k-1}\left(\mathbf{X}_k \right) &= \int {f\left( {\mathbf{X}_k|\mathbf{X}_{k-1} } \right)} {p_{k - 1}}\left( \mathbf{X}_{k-1}  \right)\delta \mathbf{X}_{k-1}, \\
\ p_k\left( \mathbf{X}_k \right) &= \frac{{{\ell_k}\left( {{Z_k}|\mathbf{X}_k} \right){p_{k|k - 1}}\left( \mathbf{X}_k \right)}}{\int {{\ell_k}\left( {{Z_k}| \mathbf{X}_k} \right)} {p_{k|k - 1}}\left( \mathbf{X}_k \right)\delta \mathbf{X}_k},
\end{align}
where $f( { \mathbf{X}_k | \mathbf{X}_{k-1} })$ denotes the transition density of trajectories and $ p_{k|k - 1}(\mathbf{X}_k)$ denotes the predicted multi-trajectory density at time $k$. Given a set of measurements $Z_k$, the density of the measurements of trajectories is denoted as ${\ell_k}\left({Z_k}|\mathbf{X}_k\right)$. Since the measurements are only based on the current target states, ${\ell_k}( {{Z_k}}|\mathbf{X}_k)$ can be also written as
\begin{equation}
{\ell_k}\left( {{Z_k}|\mathbf{X}_k} \right) = {\ell_k}\left( {{Z_k}|{\tau _k}\left( \mathbf{X}_k \right)} \right).
\end{equation}
Thus, for a single trajectory $X=(t,x^{1:i-1})$ at time $k-1$, its measurement likelihood function is expressed as $l_k(z|x^{i-1})$ and its transition density to time $k$ roots in the transition of target states from time $k-1$ to $k$, which is expressed as $f(x^{i}|x^{i-1})$.
\subsection{The TPHD and TCPHD Filters}
\subsubsection{Poisson Trajectory RFS}
{\color{black}The TPHD filter propagates the Poisson multi-trajectory density \cite{Angel2019TPHD}, with a KLD minimization after the update step.} At time $k$, the posterior multi-trajectory density $p(\cdot)$ of a Poisson RFS is given as
\begin{equation}\color{black}
p_k(\{X_1,...,X_{N^k}\})=e^{-\lambda_k}\lambda_k^n\prod_{j=1}^{N^k}\bar{p}_k(X_j),
\end{equation}
where $\bar{p}_k(\cdot)$ represents a single trajectory density and $\lambda_k\ge0$. A Poisson PDF can be characterized by its {\color{black}intensity $D_k(X) = \lambda_k\bar{p}_k(X)$ \cite{Angel2019TPHD}}. Meanwhile, the clutter RFS is Poisson with mean $\lambda_c$ and density $\bar c(\cdot)$.
\par Based on the following assumptions:
\begin{itemize}
	\item [$\bullet$]
	The trajectories at time $k$ are the union of the current new trajectories that are born independently with the PHD ${{\gamma}}( \cdot)$ of Poisson densities, and the surviving trajectories at time $k-1$ with the surviving probability ${p_{S,k}}\left(  \cdot  \right)$. The birth and the surviving RFSs are independent of each other.
	\item [$\bullet$]
	The trajectory RFS at time $k-1$ is Poisson. The clutter RFS is also Poisson and independent of measurement RFS.
\end{itemize}
	\par If at time $k-1$, the posterior PHD $D_{k - 1}(t,x^{1:i-1})$ is given, {\color{black}the prediction step of the TPHD filter for $X=(t,x^{1:i})$ is expressed as}
\begin{equation}
{D_{k|k - 1}}\left( X \right) = {\gamma _k}\left( X \right) + D_k^\zeta \left( X \right),
\end{equation}
where
\begin{align}
\label{equ-birth-PHD}{\gamma _k}\left( t,x^{1:i} \right) &= \gamma \left( {t,{x^{1:i}}} \right)\delta_1[i]\delta_k[t],\\
D_k^\zeta \left( t,x^{1:i} \right) &= {p_{S,k}}\left( x^{i-1} \right)f\left( {x^i|x^{i-1}} \right){D_{k - 1}}\left( t,x^{1:i-1} \right).\label{pr_TPHD}
\end{align}
\par It is required $t \in \{1,2,...,k-1\}$ to denote trajectories born before time $k$ in (\ref{pr_TPHD}). The latest time of a trajectory $X=(t,x^{1:i})$ is~$t+i-1$, thus the predicted PHD $D_k^\zeta$ is zero if $t+i-1 \neq k$ that indicates
the trajectory is dead, as only alive
trajectories are considered in \cite{Angel2019TPHD}. {\color{black}The notation $f(\cdot|\cdot)$ represents the transition density of the target in \eqref{pr_TPHD}.} The update step of the TPHD filter is given by
\begin{equation}
\begin{split}
{D_k}\left( X \right) =& {D_{k|k - 1}}\left( X \right){q_{D,k}\left( x^i \right)}\\
&+{D_{k|k - 1}}\left( X \right)p_{D,k}(x^i)\\
&\times\sum\limits_{z \in {Z_k}} {\frac{{{l_k}(z|x^i)}}{{{\lambda _c}\bar c (z) + \left\langle {p_{D,k}\cdot{l_k}\left( {z| \cdot } \right),D_{k|k-1}^\tau } \right\rangle }}},
\end{split}
\end{equation}
where
\begin{equation}
D_{k|k-1}^\tau \left( {x^i}\right)=\sum\limits_{t = 1}^k {\int {{D_{k|k - 1}}(t,{x^{1:k - t+1}})d{x^{1:k - t}}} },
\end{equation}
which denotes the PHD of the prior target density at time $k$ and is obtained by the marginalization for Poisson multi-trajectory densities \cite{Angel2019TPHD}. The detection probability is denoted as $p_{D,k}(\cdot)$ and the notation $q_{D,k}$ equals to $1-p_{D,k}$.
\subsubsection{IID Cluster Trajectory RFS}
The TCPHD filter considers an IID cluster multi-trajectory density \cite{Angel2019TPHD}. At time $k$, the posterior multi-trajectory density $p_k(\cdot)$ of an IID cluster RFS is given as
\begin{equation}\color{black}
p_k(\{X_1,...,X_{N^k}\})=\rho_k(n)n!\prod_{j=1}^{N^k}\bar{p}_k(X_j),
\end{equation}
where $\rho_k(\cdot)$ is the cardinality distribution and $\bar{p}_k(\cdot)$ is the single trajectory density. The PHD of posterior multi-trajectory density is given as
\begin{equation}
D_k(X) =\bar{p}_k(X)\sum_{n=0}^{\infty}n\rho_k(n).
\end{equation}
Based on the following assumptions:
\begin{itemize}
	\item [$\bullet$]
	The trajectories at time $k$ are the union of the surviving trajectories at time $k-1$ and the current new trajectories with cardinality distribution ${\rho _{\gamma ,k}}\left(  \cdot  \right)$. The birth and the surviving RFS are independent of each other.
	\item [$\bullet$]
	Both the trajectory RFS and the clutter RFS are IID cluster. The cardinality distribution of trajectory and clutter are respectively given as ${\rho_k}(\cdot)$ and ${\rho _{c,k}}(\cdot)$. The clutter RFS is independent of measurement RFS.
\end{itemize}
	\par Based on the KLD minimization, the TCPHD filter propagates both the PHD of the IID multi-trajectory densities and the cardinality distribution \cite{Angel2015KLD}. Given the posterior cardinality distribution $\rho_{k-1}(\cdot)$ and posterior PHD $D_{k-1}(t,x^{1:i-1})$ at time $k-1$, the prediction step of the TCPHD filter is obtained by
\begin{align}
{D_{k|k - 1}}\left( X \right) =& {\gamma _k}\left( X \right) + D_k^\zeta \left( X \right),\\
{\rho _{k|k - 1}}\left( n \right) = &\sum\limits_{j = 0}^n {{\rho _{\gamma ,k}}\left( {n - j} \right)} \sum\limits_{\ell  = j}^\infty  {C_j^\ell } {\rho _{k - 1}}\left( \ell  \right)\notag\\
&\times \frac{{{{\left\langle {{p_{S,k}},D_{k - 1}^\tau } \right\rangle }^j}{{\left\langle {1 - {p_{S,k}},D_{k - 1}^\tau } \right\rangle }^{\ell  - j}}}}{{{{\left\langle {1,D_{k - 1}^\tau } \right\rangle }^\ell }}}.
\end{align}
The update step of the TCPHD filter is expressed as
\begin{align}
{\rho _k}\left( n \right) = &\frac{{\Upsilon _k^0\left[ {D_{k|k-1}^\tau ;{z_k}} \right](n){\rho _{k|k - 1}}\left( n \right)}}{{\left\langle {\Upsilon _k^0\left[ {D_{k|k-1}^\tau ;{z_k}} \right],{\rho _{k|k - 1}}} \right\rangle }},\\
{D_k}\left( X \right) =& {D_{k|k - 1}}\left( X \right){q_{D,k}\left( x^i \right)} \notag\\
&\times\frac{{\left\langle {\Upsilon _k^1\left[ {D_{k|k-1}^\tau ;{Z_k}} \right],{\rho _{k|k - 1}}} \right\rangle }}{{\left\langle {\Upsilon _k^0\left[ {D_{k|k-1}^\tau ;{Z_k}} \right],{\rho _{k|k - 1}}} \right\rangle }}\\
&+{D_{k|k - 1}}\left( X \right) {{p_{D,k}}\left( x^i \right)}\notag\\
&\times\sum\limits_{z \in {Z_k}}{ \frac{{l_k\left( {z|x^i} \right)}}{{\bar c\left( z \right)}}\frac{{\left\langle {\Upsilon _k^1\left[ {D_{k|k-1}^\tau ;{Z_k}\backslash \left\{ z \right\}} \right],{\rho _{k|k - 1}}} \right\rangle }}{{\left\langle {\Upsilon _k^0\left[ {D_{k|k-1}^\tau ;{Z_k}} \right],{\rho _{k|k - 1}}} \right\rangle }}}, \notag
\end{align}
where
\begin{align}
{\Upsilon ^u}&\left[ {D_{k|k-1}^\tau ,{Z_k}} \right]\left( n \right)\\
=&\sum\limits_{j = 0}^{\min \left( {M^k,n - u} \right)} {\left( {M^k - j} \right)!{\rho _{c,k}}} \left( {M^k - j} \right)\notag\\
&\times P_{j + u}^n \frac{{{{\left\langle {q_{D,k},D_{k|k-1}^\tau } \right\rangle }^{n - j - u}}}}{{{{\left\langle {1,D_{k|k-1}^\tau } \right\rangle }^n}}}{e_j}\left( {\Xi \left( {D_{k|k-1}^\tau ,{Z_k}} \right)} \right),\notag\\
\Xi& \left( {D_{k|k-1}^\tau ,{Z_k}} \right)\\
=&\left\{\int {{p_{D,k}}\left( x^i \right)\frac{{{l_k}\left( {z|x^i} \right)}}{{\bar c\left( z \right)}}} D_{k|k - 1}^\tau \left( x^i \right)dx^i:z \in {Z_k}\right\}.\notag
\end{align}
\par The TPHD and TCPHD filters can be effectively implemented by the Gaussian mixtures \cite{Angel2019TPHD}, which are respectively referred to as the GM-TPHD and GM-TCPHD filters. Besides, the pruning and absorption procedures are also proposed to prevent an unbounded increase of Gaussian components.
\section{Trajectory PHD and CPHD Filters with Unknown Detection Profile}
In this section, the prediction and update steps of the U-TPHD and U-TCPHD filters are elaborated. For both filters, the sequence of the unknown detection probabilities is augmented into the trajectory, which can not only learn the uneven detection profile, but also obtain the augmented target trajectories. In principle, by updating the sequence of the detection probabilities, both filters can perform more robustly in scenarios where targets are with unknown and time-varying detection probabilities than considering a single frame time.
\subsection{The Augmented State Space Model}
Let $\mathbb U_k$ represent the space of augmented trajectories and $\mathbb{D}^i$ denote {\color{black}the space of the sequence of the unknown detection probabilities,} where $\mathbb D$ denotes the space in the interval of $[0,1]$. Then, the augmented trajectory space model is defined as
\begin{align}\color{black}
\mathbb U_k  = \uplus_{(t,i)\in \mathbb{I}_k}\{t\}\times\mathbb{R}^{i\times n_x}  \times \mathbb{D}^{i}.
\end{align}
Supposing that at time $k$, there are $N^k$ augmented trajectories taking values in the state space $\mathbb U_k$, then the augmented trajectory RFS is given as
\begin{align}
	\mathbf{\tilde X}_k=&\{\tilde X_{1},...,\tilde X_{N^k}\}\in \mathcal{F}(\mathbb U_k),
\end{align}
which is a simple development of trajectory RFS. Each element of $\mathbf{\tilde X}$ denotes an augmented trajectory state and is expressed as
\begin{align}
\label{equ_trajectory}{\tilde X}=  ({X,A}) \in \mathbb U,
\end{align}
{\color{black}which consists of the trajectory $X=(t,x^{1:i}) \in \mathbb T$ and the sequence of detection probabilities $A=a^{1:i} \in \mathbb {D}^i$. On the other hand, at each time step of the augmented trajectory, we can obtain the augmented target state $\tilde x =(x,a)$ information, where $a\in\mathbb D$ denotes the detection probability and $x$ denotes the target state. The density of a single augmented trajectory is given as $\bar p(\tilde X)$ and its integral is}
{\color{black}
\begin{align}
	\int_\mathbb{U}{\bar p\left(\tilde X\right)d\tilde X}=&\sum_{(t,i)\in \mathbb{I}}\int_{\mathbb{R}^{in_x}}\int_{\mathbb{D}^{i}}\bar p\left(t,x^{1:i},a^{1:i}\right)da^{1:i}dx^{1:i}.
\end{align}
}
The density of sets of augmented trajectories is defined as the augmented multi-trajectory density. Since the detection only concerns about alive targets, the detection probability at time $k$ is presented by the state at the latest time step of the augmented part $A$
\begin{align}
	p_{D,k}(\tilde X)=a^i,
\end{align}
where $i=k-t+1$. The transition of augmented trajectory is given as $\tilde f( {\tilde X|\underline{\tilde X}})$, where $\underline{\tilde X}=(\underline{t},\underline{x}^{1:\underline{i}},\underline{a}^{1:\underline{i}})$ denotes the augmented trajectory at time $k-1$. {\color{black} In order to better explain the relationship between the detection probability sequence and trajectory, we consider the switch of detection probability is independent of the trajectory state and the transition is the first-order Markov process, then the equation is established as follows}
{\color{black}
\begin{align}\label{trans}
	\tilde f\left( {\tilde X|\underline{\tilde X}} \right) =&\tilde f\left( {t},{x}^{1:i},{a}^{1:i}|\underline{t},\underline{x}^{1:\underline{i}},\underline{a}^{1:\underline{i}} \right)\\
	=& f\left( {x^i}|{x^{i - 1},a^{i-1}} \right)g\left( {a^i|a^{i-1}} \right)\delta_{\underline{x}^{1:\underline{i}}}({x}^{1:i-1})\notag\\
	&\times\delta_{\underline{a}^{1:\underline{i}}}({a}^{1:i-1})\delta_{\underline{t}}[{t}]\delta_{\underline{i}+1}[{i}],\notag
\end{align}
 where $f(\cdot|\cdot)$ and $g(\cdot|\cdot)$ represents the transition density of target and detection probability from time $k-1$ to $k$.}
\par {\color{black}Given the augmented trajectory $\tilde X$, the PHD of the augmented multi-trajectory density at time $k$ is denoted as $D_k(\tilde X)$ and the surviving probability of the augmented trajectory is given as}
\begin{align}
{p_{S,k}}\left( {\tilde X} \right) = {p_{S,k}}\left( {X,A} \right) = {p_{S,k}}\left( x^i \right).
\end{align}
In this paper, the measurements are only concerned about kinematic states of targets, so the measurement likelihood $l_k (z|\tilde X)$ can be simplified to
\begin{align}
{l_k}\left( {z|\tilde X} \right) = {l_k}\left( {z|x^i} \right).
\end{align}
\subsection{The U-TPHD Filter}
In this section, the recursion of the U-TPHD filter is derived in detail, which predicts and updates both the trajectories and histories of the detection profile rather than target states and detection profile only at the latest frame time \cite{Mahler2011UPHD}. {\color{black}In order to solve the problem that updated posterior augmented multi-trajectory densities are no longer Poisson, the U-TPHD filter finds the best Poisson approximation through the KLD minimization \cite{Angel2015KLD}.} Based on this theory, the U-TPHD filter propagates the PHD of a Poisson augmented multi-trajectory density through recursions. The prediction and update steps are given in Propositions \ref{UTPHD_Pr} and \ref{UTPHD_up}.
\par The recursion of the U-TPHD filter is following the routine of:
\par\emph{Assumption 1:} The trajectories are the superposition of alive trajectories at the last time and new births at the current time, which are independent of each other. The PHD of the born trajectory with the augmented
part is given as $\gamma ( {\tilde X})$ and the birth model is assumed as known.
\par\emph{Assumption 2:} The clutter RFS is Poisson with mean ${\lambda _c}$ and density $\bar c\left( \cdot \right)$. The clutter RFS is independent of the measurement RFS.
\par\emph{Assumption 3:} Each trajectory generates measurements independently of each other.
\par\emph{Assumption 4:} The prior and posterior augmented multi-trajectory densities are Poisson.
\begin{Pro}\label{UTPHD_Pr}
{\color{black}If at time $k-1$, the posterior PHD $D_{k-1}(t,x^{1:i-1},a^{1:i-1})$ is given, then the predicted PHD $D_{k|k-1}(\cdot)$ for $\tilde X=(t,x^{1:i},a^{1:i})$ is given as}
\end{Pro}
\begin{align}\label{equ_UTPHD_pr_total}
{D_{k|k - 1}}\left( {\tilde X} \right)= {\gamma _k}\left( {\tilde X} \right) + D_k^\zeta \left( {\tilde X} \right),
\end{align}
where
\begin{align}
{\gamma _k}\left( t,x^{1:i},a^{1:i} \right) =&  \gamma \left( {t,{x^{1:i}},a^{1:i}} \right)\delta_{1}[i]\delta_{k}[t],\\
\label{equ_UTPHD_pr}D_k^\zeta \left(  t,x^{1:i},a^{1:i} \right) =&  {p_{S,k}}\left( {{x^{i-1}}} \right)f\left( {x^i}|{x^{i - 1}},a^{i-1} \right) g\left( {a^i|a^{i-1}}\right)\notag\\
&\times{D_{k - 1}}\left( t,x^{1:i - 1},a^{1:i - 1} \right).
\end{align}
\par In Proposition \ref{UTPHD_Pr}, it is required $t \in \{1,2,...,k-1\}$ to denote trajectories born before time $k$. Besides, the equation $t + i - 1 = k$ needs to be satisfied in (\ref{equ_UTPHD_pr}), since only alive trajectories are considered. {\color{black} The proof of Proposition \ref{UTPHD_Pr} can be found in Appendix A.}
The proof of the KLD minimization is omitted, since the trajectory augmented with the detection probability sequence is a simple extension of the classical trajectory state, and the proof of the latter can be found in \cite{Angel2019TPHD}.
\begin{Rem}
\rm	The robust PHD filter \cite{Mahler2011UPHD} only considers the target state and corresponding detection probability at the current time. {\color{black}In contrast, in Proposition \ref{UTPHD_Pr}, the past states of the trajectories and detection profile sequence are kept in the U-TPHD filter.} The condition also applies to the prediction step in the U-TCPHD filter following.
\end{Rem}
\begin{Pro}\label{UTPHD_up}
	If at time $k$, the predicted PHD $D_{k|k-1}(\tilde X)$ is given, then the posterior PHD $D_{k}(\tilde X)$ is given by
\end{Pro}
\begin{align}
\label{equ_UTPHD_up}{D_k}&\left( t,x^{1:i},a^{1:i} \right) \\
=& {D_{k|k - 1}}\left( t,x^{1:i},a^{1:i} \right)(1-a^i) + {D_{k|k - 1}}\left( t,x^{1:i},a^{1:i} \right)a^i\notag\\
&\times\sum\limits_{z \in {Z_k}} {\frac{l_k(z|x^i)}{\lambda_c\bar c (z) +  \iint a^i\cdot l_k(z|x^i)D^\tau_{k|k-1}(x^i,a^i)da^idx^i }},\notag
\end{align}
where
\begin{align}
D_{k|k-1}^\tau(x^i,a^i)=&\sum\limits_{t = 1}^k\iint {D_{k|k - 1}(t,x^{1:k-t+1},a^{1:k-t+1})}\notag\\
\label{equ_D}&{\times dx^{1:k-t}da^{1:k-t}},
\end{align}
with $i=k-t+1$.
\par In Proposition \ref{UTPHD_up}, the PHD of the prior density of augmented targets at time $k$ is expressed as $D_{k|k-1}^\tau (x^i,a^i)$, which is obtained by the marginalization for augmented multi-trajectory densities. Since we focus on the association between the latest state of augmented trajectory and the measurements at the current time. The marginalization step is also applied to the U-TCPHD filter following. The proof of Proposition \ref{UTPHD_up} is detailed in Appendix B.
\begin{Rem}
Different from the TPHD filter \cite{Angel2019TPHD}, which only considers the PHD of trajectory, the updated PHD in the U-TPHD filter contains information about both the trajectory $X=(t,x^{1:i})$ and {\color{black}detection profile sequence $A=a^{1:i}$. The detection profile sequence $A$ and the trajectory $X$ are mutually coupled, if we obtain a better estimation of the detection profile sequence, the trajectory estimation will also be influenced.}
\end{Rem}
\begin{Rem}
	{\color{black}The sequence of unknown detection probabilities can be used as a malleable theoretical basement for other feasible variations and implementations. Its single frame variation is only one of the possible approximations. Similarly, other variations to realize the theory of sequence of augmented parts can also be devised. For example, one can replace the unknown detection probability variable by target motion model variable \cite{Pasha2009JMS, JMS-TPHD} to gain the multi-model extension of the method for maneuvering target tracking.}
\end{Rem}
\subsection{The U-TCPHD Filter}
In this section, the recursion step of the U-TCPHD filter is derived in detail. The U-TCPHD propagates the augmented multi-trajectory density of an IID cluster RFS. {\color{black}In the prediction and update steps, the U-TCPHD
filter uses the KLD minimization to obtain the best IID cluster
approximation \cite{Angel2015KLD}}. Based on this theory, it propagates the PHD of an IID cluster augmented multi-trajectory density and the cardinality distribution. The following propositions show recursion steps of the U-TCPHD filter.
\par Based on Assumptions 1--3, the recursion of the U-TCPHD filter is also following the routine of:
\par\emph{Assumption 5:} The clutter RFS is IID cluster and independent
	of the measurement RFS, with the cardinality distribution $\rho_{c,k}$. Besides, the cardinality distribution of new birth trajectories is given as ${\rho _{\gamma ,k}}$.
\par\emph{Assumption 6:} Both the prior and posterior augmented multi-trajectory densities are IID cluster through the KLD minimization.
\begin{Pro}\label{UTCPHD_Pr}
	If at time $k-1$, the posterior PHD $D_{k-1}({\tilde X})$ and posterior cardinality distribution $\rho_{k-1}$ are given, then the predicted PHD $D_{k|k-1}(\tilde X)$ and predicted cardinality distribution $\rho_{k|k-1}$ are given as
\end{Pro}
\begin{align}
\label{equ_UTCPHD_pr1}&{D_{k|k - 1}}\left( {\tilde X} \right)={\gamma _k}\left( {\tilde X} \right) + D_k^\zeta \left( {\tilde X} \right),\\
\label{equ_UTCPHD_pr2}&{\rho _{k|k - 1}}\left( n \right)= \sum\limits_{j = 0}^n {{\rho _{\gamma ,k}}\left( {n - j} \right)} \sum\limits_{\ell  = j}^\infty  {C_j^\ell } {\rho _{k - 1}}\left( \ell  \right)\\
&\times{{\left[\iint \left(1-p_{S,k}({x}^{i-1})\right)D^\tau_{k-1}({x}^{i-1},{a}^{i-1})da^{i-1}d{x}^{i-1} \right]^{\ell  - j}}}\notag\\
&\times\frac{\left[\iint p_{S,k}({x}^{i-1})D^\tau_{k-1}({x}^{i-1},{a}^{i-1})d{a}^{i-1}d{x}^{i-1} \right]^j}{\left[\iint D^\tau_{k-1}({x}^{i-1},{a}^{i-1})d{a}^{i-1}d{x}^{i-1} \right]^\ell }.\notag
\end{align}
where
\begin{align}
	D_{k-1}^\tau({x}^{i-1},{a}^{i-1})=&\sum\limits_{{t} = 1}^{k-1}\iint {D_{k - 1}({t},{x}^{1:i-1},{a}^{1:i-1})}\notag\\
	&{\times d{x}^{1:i-2}d{a}^{1:i-2}},
\end{align}
\begin{Pro}\label{UTCPHD_up}
	If at time $k$, the predicted PHD $D_{k|k-1}(\tilde X)$ and predicted cardinality distribution $\rho_{k|k-1}$ are given, where $\tilde X=(t,x^{1:i},a^{1:i})$, then the posterior PHD $D_{k}$ and posterior cardinality distribution $\rho_{k}$ are given as
\end{Pro}
\begin{align}
\label{equ_UTCPHD_up1}{D_k}\left( t,x^{1:i},a^{1:i}\right)=& {D_{k|k - 1}}\left( t,x^{1:i},a^{1:i} \right) ({1-a^i})\\
&\times\frac{{\left\langle {\Upsilon _k^1\left[ {D_{k|k-1}^\tau ;{Z_k}} \right],{\rho _{k|k - 1}}} \right\rangle }}{{\left\langle {\Upsilon _k^0\left[ {D_{k|k-1}^\tau ;{Z_k}} \right],{\rho _{k|k - 1}}} \right\rangle}}\notag\\
&+ {D_{k|k - 1}}\left( t,x^{1:i},a^{1:i} \right){a^i}\sum\limits_{z \in {Z_k}} {\frac{{l_k\left( {z|x^i} \right)}}{{\bar c\left( z \right)}}}\notag\\
&\times\frac{{\left\langle {\Upsilon _k^1\left[ {D_{k|k-1}^\tau ;{Z_k}\backslash \left\{ z \right\}} \right],{\rho _{k|k - 1}}} \right\rangle }}{{\left\langle {\Upsilon _k^0\left[ {D_{k|k-1}^\tau ;{Z_k}} \right],{\rho _{k|k - 1}}} \right\rangle }},\notag\\
\label{equ_UTCPHD_up2}{\rho _k}\left( n \right) =& \frac{{\Upsilon _k^0\left[ {D_{k|k-1}^\tau ;{Z_k}} \right](n){\rho _{k|k - 1}}\left( n \right)}}{{\left\langle {\Upsilon _k^0\left[ {D_{k|k-1}^\tau ;{Z_k}} \right],{\rho _{k|k - 1}}} \right\rangle }},
\end{align}
where
\begin{align}
{\Upsilon_k ^u}&\left[ {D_{k|k-1}^\tau ,{Z_k}} \right]\left( n \right)\notag\\
=& \sum\limits_{j = 0}^{\min \left( {\left| {{Z_k}} \right|,n - u} \right)} {\left( {\left| {{Z_k}} \right| - j} \right)!{\rho _{c,k}}} \left( {\left| {{Z_k}} \right| - j} \right)\notag\\
&\times P_{j + u}^n{e_j}\left( {\Xi \left( {D_{k|k-1}^\tau ,{Z_k}} \right)} \right)\\
&\times\frac{\left[\iint (1-a^i)D^\tau_{k|k-1}(x^i,a^i)da^idx^i \right]^{n - j - u}}{\left[\iint D^\tau_{k|k-1}(x^i,a^i)da^idx^i \right]^n},\notag\\
\Xi &\left( {D_{k|k-1}^\tau ,{Z_k}} \right)\\
 =& \left\{ {\iint {a^i\cdot\frac{{{l_k}\left( {z|x^i} \right)}}{{\bar c\left( z \right)}}} D_{k|k-1}^\tau \left( {x^i,a^i}\right)da^idx^i :z \in {Z_k}} \right\}\notag.
\end{align}
In Proposition 3, the notation $D_{k - 1}^\tau$ represents the PHD of the posterior density of augmented targets at time $k-1$, which can be obtained by \eqref{equ_D}.
In Proposition 4, the updated cardinality $\rho_k(\cdot)$ incorporates the clutter cardinality, the measurement set, the prior PHD and predicted cardinality distribution. It should be noted that at each recursion step, the cardinality distribution for a set of augmented trajectories is the
same as that for a set of targets. Therefore, the update of cardinality distribution of the U-TCPHD filter and the classic CPHD filter \cite{Vo2007CPHD} enjoy similar principles. The proof of Proposition 4 can be found in Appendix C.
\subsection{Possible Extension}	{\color{black}Similarly, the U-TPHD filter can be also extended to consider the unknown clutter rate \cite{Mahler2011UPHD}. In this situation, the hybrid augmented trajectory space model is given as $\mathbb{Y}= \mathbb{U} \uplus(\mathbb{C} \times \mathbb{D})$, where $\mathbb{C}$ and $\mathbb{D}$ denote the state space of clutter $c$ and its detection probability $o$, respectively. It is assumed that trajectories and clutter generators are statistically independent. The integral function of the density $\bar p(\cdot)$ of a single hybrid augmented trajectory $\tilde X_c \in \mathbb{Y}$ is given as
	\begin{align}
		\int_\mathbb{Y}{\bar p\left(\tilde X_c\right)d\tilde X_c}=&\sum_{(t,i)\in \mathbb{I}}\int_{\mathbb{R}^{in_x}}\int_{\mathbb{D}^{i}}\bar p\left(t,x^{1:i},a^{1:i}\right)da^{1:i}dx^{1:i}\notag\\
		&+\int_{\mathbb{C}}\int_{\mathbb{D}}\bar p\left(c,o\right)dcdo.
	\end{align}
Let $D_{k}(c,o)$ denote the PHD of the clutter generators, $p^c_{S,k}$ denote survival probability and $\gamma_{k}(c,o)$ denote PHD of the birth clutter from clutter generators at time $k$. Enjoying the same principle and assumptions of \cite{Mahler2011UPHD}, the recursion of the new U-TPHD filter is given as
	\begin{align}
		{D_{k|k - 1}}\left( {\tilde X} \right)=& {\gamma _k}\left( {\tilde X} \right) + D_k^\zeta \left( {\tilde X} \right),\\
		{D_{k|k - 1}}\left( c,o \right)=& {\gamma _k}\left( c,o \right) + p^c_{S,k}D_{k-1}\left( c,o \right),\\
		{D_k}\left( {\tilde X} \right) =& {D_{k|k - 1}}\left( {\tilde X} \right)(1-a^i)\notag\\
		& + {D_{k|k - 1}}\left( {\tilde X} \right)a^i\sum\limits_{z \in {Z_k}} {\frac{l_k(z|x^i)}{\Theta_k\left[z,c,o,\tilde x\right]}},\\
		{D_k}\left( c,o \right) =& {D_{k|k - 1}}\left( c,o \right)(1-o)\notag\\
		& + {D_{k|k - 1}}\left( c,o \right)o\sum\limits_{z \in {Z_k}}{\frac{\bar c_k(z)}{\Theta_k\left[z,c,o,\tilde x\right]}},
	\end{align}
where
	\begin{align}
		\Theta_k\left[z,c,o,\tilde x\right]=&\iint o\cdot \bar c(z)D_{k|k-1}(c,o)dcdo\\
		&+\iint a^i\cdot l_k(z|x^i)D^\tau_{k|k-1}(x^i,a^i)da^idx^i,\notag
	\end{align}
The extension of U-TCPHD filter also propagates the cardinality distribution of the hybrid augmented trajectory. The specific derivation of the U-TCPHD filter considering both unknown clutter rate and detection probability is a similar development to U-TPHD filter. In this paper, we focus on the derivation of unknown detection probability, hence their implementation methods considering clutter rate are omitted in this paper. Further details about unknown clutter rate can be found in \cite{Mahler2011UPHD}.}

\section{Beta-Gaussian Mixture Implementation for the U-TPHD and U-TCPHD Filters}
In this section, a closed form implementation for the U-TPHD and U-TCPHD filters immune to the unknown detection
profile is derived. In Section IV-A, the reason of simplifying both filters to consider the detection profile only for a single frame time is discussed. In Section IV-B, the Beta-Gaussian mixture implementation \cite{Mahler2011UPHD} is presented for the U-TPHD and U-TCPHD filters, which are referred to as the BG-U-TPHD and BG-U-TCPHD filters. In Section IV-C, the $L$-scan approximations of the BG-U-TPHD and BG-U-TCPHD filters are presented to reduce the computational burden. The estimation step is given in Section IV-D.

\subsection{Only Current Detection Profile}
It can be seen from the Propositions 1--4 that the U-TPHD and U-TCPHD filters update both the detection profile sequence and trajectory in recursions. In this section, we advocate for only considering the unknown detection profile at the current time for simplicity. 


Regarding the algorithmic efficiency, the computational cost rises significantly with the increasing length of the trajectory and {\color{black}the sequence of detection probabilities}. For simplicity, we only consider $P$ discrete probability values to describe the detection probability distribution, which are the uniform grids of interval $[0,1]$. Then, as indicated by (\ref{equ_UTPHD_pr}), for a certain survival trajectory $X=(t,x^{1:i-1})$ at time $k-1$, its historical detection probability space contains $P ^{i-1}$ components, and the number of components will increase to $P ^ i$ at time $k$. {\color{black} In other words, the computational burden of a single trajectory with length $l$ is $\mathcal{O}(P^l)$, which is not acceptable in the implementation.}

Therefore, in consideration of the algorithmic efficiency, we propose to consider the effect of the unknown detection probability of the current time for the implementations of the U-TPHD and U-TCPHD filters. {\color{black}In \cite{Mahler2011UPHD}, the Beta-Gaussian mixture provides an efficient method to directly describe the unknown detection probability, and this method can be also adopted for the U-TPHD and U-TCPHD filters.} 
{\color{black}For similarity, the unknown detection probability only at the current time is considered in the implementation. Based on this condition, the transition density $f(\cdot|\cdot)$ of targets in \eqref{trans} is assumed to be independent of the detection probability. Thus, the augmented trajectory of (\ref{equ_trajectory}) becomes to $\tilde X=(t,x^{1:i},a^i)$ and the prediction of augmented trajectory PHD (\ref{equ_UTPHD_pr_total}) can be written as}
\begin{align}\label{equ_Dtau}
	{D_{k|k - 1}}&\left( t,x^{1:i},a^i \right)\notag\\
	=& {\gamma}\left( t,x^{1:i},a^i \right)\delta_1[i]\delta_k[t]+ p_{S,k}\left( {{x^{i - 1}}} \right)f\left( {{x^i}|{x^{i - 1}}} \right)\notag\\
&\times\int g\left( {a^i|{a}^{i-1}} \right){D_{k - 1}}\left( {{t},{{x}^{1:i - 1}},{a}^{i-1}} \right)d{a}^{i-1}.
\end{align}k
Different from (\ref{equ_UTPHD_pr}) and (\ref{equ_UTCPHD_pr1}), the past states of the detection probability sequence are not retained in the prediction step and both filters only update the detection probability at the current time. After marginalization, the PHD of the augmented target in (\ref{equ_D}) can be computed as
\begin{align}
	D_{k|k-1}^\tau (x^i,a^i) =\sum\limits_{t = 1}^k {\int {{D_{k|k - 1}}(t,{x^{1:k - t+1}},a^i)d{x^{1:k - t}}} },
\end{align}
with $i=k-t+1$. The change of (\ref{equ_D}) applies to (\ref{equ_UTPHD_up}), (\ref{equ_UTCPHD_pr2}), (\ref{equ_UTCPHD_up1}) and (\ref{equ_UTCPHD_up2}).
\subsection{The BG-U-TPHD and BG-U-TCPHD Filters}
In this section, the closed-form implementations are presented for the U-TPHD and U-TCPHD filters using the Beta-Gaussian mixtures \cite{Mahler2011UPHD}. At time $k$, the Gaussian density of the trajectory born at time $t$ of length $i$ is denoted as~\cite{Angel2019TPHD}
 \begin{align}
	{\cal N}( {t,x^{1:i};t^k,\widehat{m}^k,\widehat{P}^k})={\cal N}( {x^{1:i};\widehat{m}^k,\widehat{P}^k})\delta_{[t^k]}[t]\delta_{[i_k]}[i].
\end{align}
where $\widehat{m}^k \in \mathbb{R}^{i{n_x}}$ and $\widehat{P}^k \in \mathbb{R}^{i{n_x}\times i{n_x}}$ denote, respectively, the mean and covariance. The term $t^k=k-i_k+1$ denotes the trajectory born time with $i_k=\text{dim}(\widehat{m}^k/n_x)$. For a matrix $V$, the notation ${V_{\left[ {n:m,s:t} \right]}}$ represents the submatrix of $V$ for rows from time steps $n$ to $m$ and columns from time steps $s$ to $t$. The notation $V_{[n:m]}$ is used to present the submatrix of $V$ for rows from time steps $n$ to $m$. Besides, the notations $V_{[n,s:t]}$ and $V_{[n]}$ represent $V_{[n:n,s:t]}$ and $V_{[n:n]}$, respectively.
\par {\color{black}The PDF of a Beta distribution is given as}
\begin{equation}
	\beta \left( { y ;u,v} \right)=\frac{y^{u-1}(1-y)^{v-1}}{\int_0^1{y^{u-1}(1-y)^{v-1}dy}},
\end{equation}
{\color{black}where the denominator $\int_0^1{y^{u-1}(1-y)^{v-1}dy}$ denotes the Beta function $B(u,v)$ and $u > 1, v > 1$. 
Some properties of the Beta distribution are summarized as follows \cite{Mahler2011UPHD},}
\begin{align}
	(1-y)\beta \left( { y ;u,v} \right) =& \frac{v}{u+v}\beta \left( { y ;u,v+1} \right),\\
	y\beta \left( { y ;u,v} \right) =& \frac{u}{u+v}\beta \left( { y;u+1,v} \right).
\end{align}
{\color{black}The transition density for the detection probability in (\ref{equ_Dtau}) is given as}
\begin{align}
	g\left( {a^i|a^{i-1}} \right) =&	\beta \left( { a ;u^{k|k-1},v^{k|k-1}} \right).
\end{align}
{\color{black}where
\begin{align}
	a^i=a=&\frac{{u^{k|k - 1}}}{{u^{k|k - 1} + v^{k - 1}}},\\
	\label{equ_u}u^{k|k - 1}=& \left( {\frac{{\mu^{k|k - 1}\left( {1 - \mu^{k|k - 1}} \right)}}{{{{\left[ {\sigma^{k|k - 1}} \right]}^2}}} - 1} \right)\mu ^{k|k - 1},\\
	\label{equ_v}v^{k|k - 1} =& \left( {\frac{{\mu^{k|k - 1}\left( {1 - \mu ^{k|k - 1}} \right)}}{{{{\left[ {\sigma^{k|k - 1}} \right]}^2}}} - 1} \right)\left( {1 - \mu^{k|k - 1}} \right),\\
	\mu^{k|k - 1} =& \frac{{u^{k - 1}}}{{u^{k - 1} + v^{k - 1}}}=a^{i-1},\\
	{\left[ {\sigma^{k|k - 1}} \right]^2} =& |{k_\beta }|\frac{{u^{k - 1}v^{k - 1}}}{{{{\left( {u^{k - 1} + v^{k - 1}} \right)}^2}\left( {u^{k - 1} + v^{k - 1} + 1} \right)}},
\end{align}
with $|k_\beta|$ is a constant slightly bigger than 1.}
\subsubsection{The BG-U-TPHD Filter} there are some assumptions given as follows.
\par \emph{Assumption 7:} The target kinematics and observation models are given as the linear Gaussian model \cite{Angel2019TPHD}
	\begin{align}
	f\left( {{x^i}|{x^{i - 1}}} \right)=&{\cal N}\left( {{x^i};F{x^{i - 1}},Q} \right),\\
	l\left( {z|x^i} \right)=&{\cal N}\left( {z;Hx^i,R} \right),
	\end{align}
where $F \in \mathbb{R}^{n_x \times n_x}$ denotes the state transition matrix, $Q\in \mathbb{R}^{n_x \times n_x}$ is the process noise covariance,
$H \in \mathbb{R}^{n_z \times n_x}$ is the observation matrix and $R \in \mathbb{R}^{n_z \times n_z}$
denotes the observation noise covariance.
\par \emph{Assumption 8:} To simplify this model, the surviving probability is taken as a constant.
\par \emph{Assumption 9:} The PHD of birth density is given as
	\begin{equation}
	{\gamma _k}\left( {\tilde X} \right) = \sum\limits_{j = 1}^{J_\gamma ^k} {\omega _{\gamma ,j}^k\beta \left( {a;u_{\gamma ,j}^k,v_{\gamma ,j}^k} \right){\cal N}\left( {X;k,\widehat{m}_{\gamma ,j}^k,\widehat{P}_{\gamma ,j}^k} \right)},
	\end{equation}
	where $J_\gamma ^k$ denotes the number of birth trajectories. For the $j$-{th} birth component at time $k$, $\omega_{\gamma,j}^k$ represents the weight. The mean and covariance of the Gaussian density of the trajectory born at time $k$ are expressed as $\widehat{m}_{\gamma,j}^k \in \mathbb{R}^{n_x}$ and $\widehat{P}_{\gamma,j}^k\in \mathbb{R}^{n_x \times n_x}$, respectively. The terms $u_{\gamma ,j}^k$ and $v_{\gamma ,j}^k$ denote the corresponding factors of the Beta distribution.
\begin{Pro}\label{BGUTPHD_Pr}
If at time $k-1$, the posterior PHD $D_{k-1}$ is given and $D_{k-1}$ is a Beta-Gaussian mixture of the form
\begin{align}
{D_{k - 1}}\left( {\tilde X} \right) =& \sum\limits_{j = 1}^{{J^{k - 1}}} {\omega _j^{k - 1}\beta \left( {a;u_j^{k - 1},v_j^{k - 1}} \right)}\notag\\
&\times{\cal N}\left(X;t^{k-1}_j,\widehat{m}_j^{k-1},\widehat{P}_j^{k-1} \right)
\end{align}
where, at time $k-1$, the $j$-{th} trajectory has length $i_j^{k-1}=k-t_j^{k-1}$. The mean and covariance of the Gaussian density are given as $\widehat m_j^{k-1}\in \mathbb{R}^{i_j^{k-1}n_x}$ and $\widehat P_j^{k-1}\in \mathbb{R}^{i_j^{k-1}n_x \times i_j^{k-1}n_x }$, respectively. Thus, the prior PHD $D_{k|k-1}$ is given as
\end{Pro}
\begin{equation}
\begin{split}
{D_{k|k - 1}}\left( {\tilde X} \right) =& \sum\limits_{j = 1}^{J_\gamma ^k} {\omega _{\gamma ,j}^k\beta \left( {a;u_{\gamma ,j}^k,v_{\gamma ,j}^k} \right)}\\
&{\times{\cal N}\left( {X;k,\widehat{m}_{\gamma ,j}^k,\widehat{P}_{\gamma ,j}^k} \right)} \\
&+ {p_s}\sum\limits_{j = 1}^{{J^{k - 1}}} {\omega _j^{k - 1}\beta \left( {a;u_{S,j}^{k|k - 1},v_{S,j}^{k|k - 1}} \right)}\\
&{\times{\cal N}\left( {X;t^{k|k-1}_{S,j},\widehat{m} _{S,j}^{k|k-1},\widehat{P} _{S,j}^{k|k - 1}} \right)},
\end{split}
\end{equation}
where
\begin{align}
\widehat m_{S,j}^{k|k - 1} =& \left[ {\left[\widehat m_j^{k - 1}\right]^\top,\left[F \cdot \widehat{m}_{j,[k-1]}^{k - 1}\right]^\top} \right]^\top,\\
\widehat P_{S,j}^{k|k - 1} =& \left[ {\begin{array}{*{20}{c}}
{\widehat P_j^{k - 1}}&P_1\\
P_1^\top&P_2
\end{array}} \right],\\
P_1=&{\widehat P_{j,\left[ {t^{k - 1}_j:k - 1,k - 1} \right]}^{k - 1}{F^\top}},\\
P_2=&{F\widehat{P}_{j,[k-1,k-1]}^{k - 1}F^\top + Q}.
\end{align}
\par The prediction of each Beta-Gaussian component is obtained by the prediction of the Beta part (denoting the detection probability) multiplied by the prediction of the Gaussian part (denoting the trajectory). Compared to the prediction step of the GM-TPHD filter \cite{Angel2019TPHD}, the prediction of trajectory in the BG-U-TPHD filter still roots in the transition of the target kinematic state, while the prediction of detection probability is completely governed by Beta densities.
The $t^{k|k-1}_S$ is the birth time of the surviving trajectory.
\begin{Pro}\label{BGUTPHD_Up}
	If at time $k$, the prior PHD $D_{k|k-1}$
	is given and $D_{k|k-1}$ is a Beta-Gaussian mixture of the form
	\begin{align}
{D_{k|k - 1}}\left( {\tilde X} \right) =& \sum\limits_{j = 1}^{{J^{k|k - 1}}} {\omega _j^{k|k - 1}\beta \left( {a;u_j^{k|k - 1},v_j^{k|k - 1}} \right)}\notag\\
&{\times{\cal N}\left( {X;t_j^{k|k - 1},\widehat m_j^{k|k - 1},\widehat P_j^{k|k - 1}} \right)}.
	\end{align}
Then, given a measurement set $Z_k$, the posterior PHD $D_k$ is given as
\end{Pro}
\begin{align}
{D_k}\left( {\tilde X} \right)=& \sum\limits_{j = 1}^{{J^{k|k - 1}}} {\omega _j^{v,k}\beta } \left( {a;u_j^{k|k - 1},v_j^{k|k - 1} + 1} \right)\\
&\times{\cal N}\left( {X;t_j^{k|k - 1},\widehat m_j^{k|k - 1},\widehat P_j^{k|k - 1}} \right)\notag\\
&+ \sum\limits_{z \in {Z_k}} {\sum\limits_{j = 1}^{{J^{k|k - 1}}} {\omega _j^{u,k}(z)\beta } \left( {a;u_j^{k|k - 1} + 1,v_j^{k|k - 1}} \right)}\notag\\
&\times{{\cal N}\left( {X;t_j^{k},\widehat m_j^k\left( z \right),\widehat P_j^k} \right)},\notag
\end{align}
where
\begin{align}
\omega _j^{v,k} =& \omega _j^{k|k - 1}\frac{{B\left( {u_j^{k|k - 1},v_j^{k|k - 1} + 1} \right)}}{{B\left( {u_j^{k|k - 1},v_j^{k|k - 1}} \right)}},\\
\omega _j^{u,k} =& \omega _j^{k|k - 1}\frac{{B\left( {u_j^{k|k - 1}+1,v_j^{k|k - 1}} \right)}}{{B\left( {u_j^{k|k - 1},v_j^{k|k - 1}} \right)}}\\
&\times\frac{{{q_j}(z)}}{{{\lambda _c}\bar c \left( z \right) + \sum\limits_{l = 1}^{{J_{k|k - 1}}} {\frac{{u_l^{k|k - 1}}}{{u_l^{k|k - 1} + v_l^{k|k - 1}}}\omega _l^{k|k - 1}{q_l}(z)} }},\notag\\
{\bar z _j} =& {H}\widehat{m}_{j,[k]}^{k|k - 1},\\
{S_j} =& {H}\widehat P^{k|k-1}_{j,[k,k]}H^\top + R,\\
{q_j}\left( z \right) =& {\cal N}\left( {z;{{\bar z }_j},{S_j}} \right),\\
\widehat{m}_j^k\left( z \right) =& \widehat{m}_j^{k|k - 1} + {K_j}\left( {z - {{\bar z}_j}} \right),\\
\widehat{P} _j^k =& \widehat{P}_j^{k|k - 1} - {K_j}{H}\widehat{P}_{j,[ {k,t^{k|k - 1}_j:k}]}^{k|k - 1},\\
{K_j} =& \widehat{P} _{j,[ {t^{k|k - 1}_j:k,k} ]}^{k|k - 1}H^\top S_j^{ - 1}.
\end{align}
\par The BG-U-TPHD filter also adopts the Kalman filter in the update step, but aims at the whole trajectory. It not only updates the estimation of the target state at the current time, but also smooths the estimation of the previous states. Different from the GM-TPHD filter \cite{Angel2019TPHD}, the updated detection probability can also be obtained by the number of measurements associated with trajectories. 
\subsubsection{The BG-U-TCPHD Filter} based on Assumptions 5--9, the BG-U-TCPHD filtering recursion is obtained by the following propositions
\begin{Pro}\label{BGUTCPHD_Pr}
	If at time $k-1$ , the posterior PHD $D_{k-1}$
	and posterior cardinality distribution $\rho_{k-1}$ are given and $D_{k-1}$ is a Beta-Gaussian mixture of the form
	\begin{align}
	{D_{k - 1}}\left( {\tilde X} \right) =& \sum\limits_{j = 1}^{{J^{k - 1}}} {\omega _j^{k - 1}\beta \left( {a;u_j^{k - 1},v_j^{k - 1}} \right)}\notag\\
	&\times{\cal N}\left(X;t_j^{k - 1},\widehat{m}_j^{k-1},\widehat{P}_j^{k-1} \right),
	\end{align}
	then at time $k$, the prior cardinality distribution $\rho_{k|k-1}$ and PHD $D_{k|k-1}$ are given as
\end{Pro}
\begin{align}
{\rho _{k|k - 1}}\left( n \right) =& \sum\limits_{j = 0}^n {{\rho _{\gamma ,k}}\left( {n - j} \right)} \sum\limits_{\ell  = j}^\infty  {C_j^\ell } {\rho _{k - 1}}\left( \ell  \right)\notag\\
& \times {p_{S,k}}^j{\left( {1 - {p_{S,k}}} \right)^{\ell  - j}},\\
{D_{k|k - 1}}\left( {\tilde X} \right) =& \sum\limits_{j = 1}^{J_\gamma ^k} {\omega _{\gamma ,j}^k\beta \left( {a;u_{\gamma ,j}^k,v_{\gamma ,j}^k} \right)}{{\cal N}\left( {X;k,\widehat{m}_{\gamma ,j}^k,\widehat{P}_{\gamma ,j}^k} \right)}\notag \\
&+ {p_{S,k}}\sum\limits_{j = 1}^{{J^{k - 1}}} {\omega _j^{k - 1}\beta \left( {a;u_{S,j}^{k|k - 1},v_{S,j}^{k|k - 1}} \right)}\\
&\times{{\cal N}\left( {X;t_{S,j}^{k|k - 1},\widehat{m} _{S,j}^{k|k - 1},\widehat{P}_{S,j}^{k|k - 1}} \right)}.\notag
\end{align}
\par In Proposition 7, the derivation of the prior PHD is the same as the BG-U-TPHD filter, which retains previous states of trajectories, while the BG-U-TCPHD filter also contains the prediction of cardinality distribution.
\begin{Pro}\label{BGUTCPHD_Up}
	If at time $k$, the prior PHD $D_{k|k-1}$
	is given and $D_{k|k-1}$ is a Beta-Gaussian mixture of the form
	\begin{align}
	{D_{k|k - 1}}\left( {\tilde X} \right) =& \sum\limits_{j = 1}^{{J^{k|k - 1}}} {\omega _j^{k|k - 1}\beta \left( {a;u_j^{k|k - 1},v_j^{k|k - 1}} \right)}\notag\\
	&{\times{\cal N}\left( {X;t_j^{k|k - 1},\widehat m_j^{k|k - 1},\widehat P_j^{k|k - 1}} \right)},
	\end{align}
	receiving a set of measurements $Z_k$, the cardinality distribution $\rho_{k}$ and the posterior PHD $D_{k}$ at time $k$ can be obtained as follows,
\end{Pro}
\begin{align}
{D_k}\left( {\tilde X} \right)=& \sum\limits_{j = 1}^{{J^{k|k - 1}}} {\omega _j^{v,k}\beta } \left( {a;u_j^{k|k - 1},v_j^{k|k - 1} + 1} \right)\\
&\times{\cal N}\left( {X;t_j^{k|k - 1},\widehat{m} _j^{k|k - 1},\widehat{P}_j^{k|k - 1}} \right)\notag\\
&+ \sum\limits_{z \in {Z_k}} {\sum\limits_{j = 1}^{{J^{k|k - 1}}} {\omega _j^{u,k}(z)\beta } \left( {a;u_j^{k|k - 1} + 1,v_j^{k|k - 1}} \right)}\notag\\
&{\times{\cal N}\left( {X;t_j^{k},\widehat{m}_j^k\left( z \right),\widehat{P} _j^k} \right)},\notag\\
{\rho _k}\left( n \right) =& \frac{{\Upsilon _k^0\left[ {{a^{k|k - 1}},{\omega ^{k|k - 1}},{Z_k}} \right](n){\rho _{k|k - 1}}\left( n \right)}}{{\left\langle {\Upsilon _k^0\left[ {{a^{k|k - 1}},{\omega ^{k|k - 1}},{Z_k}} \right],{\rho _{k|k - 1}}} \right\rangle }},
\end{align}
where
\begin{align}
{\Upsilon ^u}&\left[ {{a^{k|k - 1}},{\omega ^{k|k - 1}},{Z_k}} \right]\left( n \right)\notag\\
=&\sum\limits_{j = 0}^{\min \left( {\left| {{Z_k}} \right|,n - u} \right)} {\left( {\left| {{Z_k}} \right| - j} \right)!{\rho _c}} \left( {\left| {{Z_k}} \right| - j} \right)P_{j + u}^n\\
&\times \frac{{{{\left\langle {1 - {a^{k|k - 1}},{\omega ^{k|k - 1}}} \right\rangle }^{n - j - u}}}}{{{{\left\langle {1,{\omega ^{k|k - 1}}} \right\rangle }^n}}}\notag\\
&\times{e_j}\left( {{\Lambda _k}\left( {{a^{k|k - 1}},{\omega ^{k|k - 1}},{Z_k}} \right)} \right),\notag\\
{\Lambda _k}&\left( {{a^{k|k - 1}},{\omega ^{k|k - 1}},{Z_k}} \right) \notag\\
=&\left\{ {\sum\limits_{j = 1}^{{J^{k|k - 1}}} {a_j^{k|k - 1}\frac{{q\left( z \right)}}{{\bar c\left( z \right)}}} \omega _j^{k|k - 1}:z \in {Z_k}} \right\},
\end{align}
\begin{align}
{\omega ^{k|k - 1}} =& {\left[ {\omega _1^{k|k - 1},...,\omega _{{J^{k|k - 1}}}^{k|k - 1}} \right]^{\top}},\\
{a^{k|k - 1}}=& {\left[ \frac{{u_1^{k|k - 1}}}{{u_1^{k|k - 1} + v_1^{k|k - 1}}},...,\frac{{u_{J^{k|k-1}}^{k|k - 1}}}{{u_{J^{k|k-1}}^{k|k - 1} + v_{J^{k|k-1}}^{k|k - 1}}} \right]^{\top}},\\
q\left( z \right) =& {\left[ {{q_1}\left( z \right),...,{q_{{J^{k|k - 1}}}}\left( z \right)} \right]^{\mathrm T}},\\
\omega _j^{v,k} =& \omega _j^{k|k - 1}\frac{{B\left( {u_j^{k|k - 1},v_j^{k|k - 1} + 1} \right)}}{{B\left( {u_j^{k|k - 1},v_j^{k|k - 1}} \right)}}\\
&\times\frac{{\left\langle {\Upsilon _k^1\left[ {{a^{k|k - 1}},{\omega ^{k|k - 1}},{Z_k}} \right],{\rho _{k|k - 1}}} \right\rangle }}{{\left\langle {\Upsilon _k^0\left[ {{a^{k|k - 1}},{\omega ^{k|k - 1}},{Z_k}} \right],{\rho _{k|k - 1}}} \right\rangle }},\notag\\
\omega _j^{u,k} =& \omega _j^{k|k - 1}\frac{{B\left( {u_j^{k|k - 1}+1,v_j^{k|k - 1}} \right)}}{{B\left( {u_j^{k|k - 1},v_j^{k|k - 1}} \right)}}\\
&\times\frac{{\left\langle {\Upsilon _k^1\left[ {{a^{k|k - 1}},{\omega ^{k|k - 1}},{Z_k}\backslash \left\{ z \right\}} \right],{\rho _{k|k - 1}}} \right\rangle }}{{\left\langle {\Upsilon _k^0\left[ {{a^{k|k - 1}},{\omega ^{k|k - 1}},{Z_k}} \right],{\rho _{k|k - 1}}} \right\rangle }}\frac{{{q_j}(z)}}{{\bar c \left( z \right)}}.\notag
\end{align}
Proposition 8 shows that the BG-U-TCPHD filter updates not only the whole trajectory and detection probability, but also the cardinality distribution. In order to limit unbounded Beta-Gaussian components in the BG-U-TPHD and BG-U-TCPHD filters, the pruning and absorption methods of mixture components are proposed, and the detail steps are in Table I.

\par It should be noted that, the absorption is different from merging in the robust BG-CPHD filter \cite{Mahler2011UPHD}. In this paper, the absorption is that, for two closely spaced Beta-Gaussian components, only the weight and Beta distribution factors of the smaller one are added to the higher component. Because the distance of two Beta-Gaussian components are measured based on the target states at the current time, while their past trajectory states can be extremely different. 

\begin{table}[!t]
	\centering
	\caption{The Algorithm For Pruning and Absorption}
	\label{Algorithm}
	\begin{tabular}{p{8cm}}
		\toprule
		\midrule
		\textbf{Give} posterior PHD parameters~$\{\Phi_j^k\}_{j=1}^{J_k}$, which equal to $\left\{\omega_j^k,t^k_j,i_j^k,\widehat m_j^k,\widehat P_j^k,u_j^k,v_j^k \right\}^{J^k}_{j=1}$, a pruning threshold $\mathit{T_p}$, a absorption threshold $\mathit{T_a}$ and maximum allowable number of Gaussian terms $J_{max}$. Set $\ell=0$ and $\Theta=\left\{i=1,...,J^k|\omega_i^k>\mathit{T_p}\right\}$.\\
		\textbf{Loop}\\
		\par ~~$\ell=\ell+1.$\vspace{1mm}
		\par ~~$j$\,=\,$arg$$\mathop{max}\limits_{i\in \Theta}$\,$\omega_i^k$.\vspace{1mm}
		\par ~~$L=\{i\in \Theta:(m_i^k-m_j^k)^\top(P_j^k)^{-1}(m_i^k-m_j^k)\le\mathit{T_a}\}$.\vspace{1mm}
		\par~~$\bar\omega_\ell^k=\sum\nolimits_{i \in L}\omega_i^k$,\vspace{1mm}
		\par~~	$\mu _i^k = \frac{u_i^k}{u_i^k + v_i^k},$\vspace{1mm}
		\par~~$	{\left[ {\sigma _i^k} \right]^2} = \frac{{u_i^kv_i^k }}{{{{\left( {u_i^k+ v_i^k } \right)}^2}\left( {u_i^k + v_i^k + 1} \right)}},$\vspace{1mm}
		\par~~$\bar \mu^k=\frac{1}{\bar\omega_\ell^k}\sum\nolimits_{i \in L}\omega_i^k\mu_i^k,$\vspace{1mm}
		\par~~$\left[ {\bar\sigma^k} \right]^2=\frac{1}{\bar\omega_\ell^k}\sum\nolimits_{i \in L}\omega_i^k\left[ {\sigma_i^k} \right]^2,$\vspace{1mm}
		\par~~$u_\ell^k= \left( {\frac{{\bar \mu^k\left( {1 - \bar \mu^k} \right)}}{{{{\left[ {\bar\sigma^k} \right]}^2}}} - 1} \right)\bar \mu^k,$\vspace{1mm}
		\par~~$v_\ell^k = \left( {\frac{{\bar \mu^k\left( {1 - \bar \mu^k} \right)}}{{{{\left[ {\bar\sigma^k} \right]}^2}}} - 1} \right)\left( {1 - \bar \mu^k} \right),$\vspace{1mm}
		\par~~$\bar\Phi_\ell^k=\Phi_j^k$ with weight $\bar\omega_\ell^k$ and Beta distribution factors $u_\ell^k, v_\ell^k$.\vspace{1mm}
		\par~~$\Theta=\Theta\backslash L$.\vspace{1mm}\\
		\textbf{If} $\Theta=\emptyset$, \textbf{break}\\
		if $\ell>J_{max}$ then replace $\bar\Phi_\ell^k$ by the $J_{max}$ Gaussian components with largest weights.\\	
		\textbf{Output}:$\{\bar\Phi_j^k\}_{j=1}^{min\{\ell,J_{max}\}}$.\\
		\bottomrule
	\end{tabular}
\end{table}
\subsection{L-scan Approximation}
\par The $L$-scan approximation is proposed in \cite{Angel2019TPHD} to reduce the computational burden from the increasing trajectory state. This approximation is also applied to the BG-U-TPHD and BG-U-TCPHD filters, which only updates the augmented multi-trajectory density of the last $L$ time and keeps the rest unaltered. The notation $L$ is used to denote the value of the $L$-scan approximation. When the length of trajectory $i\le L$, the prediction and update steps are the same as Section IV-B. However, when $i\ge L$, the prediction step changes to
\begin{align}
\widehat{m} _j^{L,k|k - 1} =& \left[ \left[\widehat{m}_{j,\left[ {2:L} \right]}^{L,k - 1}\right]^\top,\left[F \cdot \widehat{m}_{j,[L]}^{L,k - 1}\right]^\top \right]^\top,\\
\widehat{P}_j^{L,k|k - 1} =& \left[ {\begin{array}{*{20}{c}}
	{\widehat{P}_{j,\left[ {2:L,2:L} \right]}^{L,k - 1}}&{\widehat{P}_{j,\left[ {2:L,L} \right]}^{L,k - 1}{F^\top}}\\
	{F\widehat{P} _{j,\left[ {L,2:L} \right]}^{L,k - 1}}&{F\widehat{P}_{j,[L,L]}^{L,k - 1}F^\top + Q}
	\end{array}} \right],
\end{align}
where the mean $\widehat{m}^{L,k-1}\in \mathbb{R}^{Ln_{x}}$ and the covariance $\widehat{P}^{L,k-1}\in \mathbb{R}^{Ln_{x}\times Ln_{x}}$. The update step is given by
\begin{align}
{\bar z _j} =& {H}\widehat{m}_{j,[L]}^{L,k|k - 1},\\
{S_j} =& {H}\widehat{P}_{j,[L,L]}^{L,k|k - 1}H^\top + R,\\
\widehat{m}_j^{L,k}\left( z \right) =& \widehat{m}_j^{L,k|k - 1} + {K_j}\left( {z - {{\bar z}_j}} \right),
\end{align}
\begin{align}
\widehat{P} _j^{L,k} =& \widehat{P}_j^{L,k|k - 1} - {K_j}{H}\widehat{P}_{j,\left[ {L,1:L} \right]}^{L,k|k - 1},\\
{K_j} =& \widehat{P} _{j,\left[ {1:L,L} \right]}^{L,k|k - 1}H^\top S_j^{ - 1}.
\end{align}

\par Besides, the matrix $A_j^k=\widehat{m}_{j,[t^k_j:k-L]}^{k}\in \mathbb{R}^{(i_j^k-L)n_x}$ is needed to store the trajectory outside the $L$-scan window, even if they are not updated. The intact trajectory information at the current time, consists of the contents in the matrix $A$ and the corresponding $L$-scan window, which is written as
$\widehat{m}_j^k=[[A_j^k]^\top,[\widehat m_j^{L,k}]^\top]^\top$. In other words, both filter will propagate $\widehat m^L$, $A$ and $\widehat P^L$ instead of $\widehat m$ and $\widehat P$ by using the $L$-scan approximation. Meanwhile, in pruning and absorption procedures, the $\widehat {m}_j^k$ and $\widehat {P}_j^k$ in PHD parameter $\Phi_j^k$ are also replaced. The rest algorithms of pruning and absorption procedures do not change in Table I. {\color{black}The BG-U-TPHD and BG-U-TCPHD filters are equivalent to the robust BG-PHD and BG-CPHD filters \cite{Mahler2011UPHD} but keeping trajectory information. When the $L$=1, the BG-U-TPHD and BG-U-TCPHD filters degrade into the robust BG-PHD and BG-CPHD filters.}

\subsection{Estimation of Trajectories and Detection Profile}
{\color{black}In this section, we will {\color{black}elaborate on} the estimation of a set of augmented trajectories at each time step in the BG-U-TPHD and BG-U-TCPHD filters. It should be noted that the estimation step is behind the pruning and absorption.}
For the BG-U-TPHD filter, the estimation of the number of alive trajectories at time $k$ is given as
\begin{equation}
{N^k} =  \text{round}\left({\sum\limits_{j = 1}^{{J^k}} { {\omega _j^k}} }\right).
\end{equation}
{\color{black}Then the estimated set of augmented trajectories is given as}
\begin{equation}\color{black}
\left\{ {( {{t_1},i_1^k,\widehat m_1^k},\frac{u_1^k}{u_1^k + v_1^k} ),...,( {{t_{{N^k}}},i_{{N^k}}^k,\widehat m_{{N^k}}^k},\frac{u_{N^k}^k}{u_{N^k}^k + v_{N^k}^k})} \right\},
\end{equation}
\par For the BG-U-TCPHD filter, the estimation of the number of alive trajectories at time $k$ can be obtained as
\begin{equation}
{N^k} = \text{argmax} {\rho _k}\left( \cdot \right).
\end{equation}
The estimation of the detection profile and set of trajectories for the BG-U-TCPHD filter is the same as the BG-U-TPHD filter.
\section{Simulation Results}
This section presents numerical studies for the BG-U-TPHD and BG-U-TCPHD filters. In Section V-A, we compare the BG-U-TPHD and BG-U-TCPHD filters with different $L$ and their abilities to approximate the GM-TPHD and GM-TCPHD filters with known detection probability for the uniform detection profile \cite{Angel2019TPHD}. In Section V-B, we present the performance of the BG-U-TPHD and BG-U-TCPHD filters with lower detection probabilities, as well as their performance under the condition of an uneven and time-varying detection profile. All filters in this section are based on the $L$-scan approximation.
\subsection{Scenario 1}
Ten targets are simulated inside of a two-dimensional space with the size of $[ {{\rm{ - 2000}},{\rm{2000}}}]m\times[ {{\rm{ 0}},{\rm{2000}}} ]m$ for 100 seconds. The target state matrix is given as $x = {[ {{p_x},{p_y},{{\dot p}_x},{{\dot p}_y}} ]^\top}$ {\color{black}including the position (with unit: $m$) and velocity information (with unit: $m/s$).} The observation matrix $z = {[ {{z_x},{z_y}}]^\top}$ includes the position information. The single target transition model is given as
\begin{align*}
F =& \left[ {\begin{array}{*{20}{c}}
	{{I_2}}&{{I_2}\delta t}\\
	{{0_2}}&{{I_2}}
	\end{array}} \right] \qquad	Q = \sigma _v^2\left[ {\begin{array}{*{20}{c}}
{\frac{{\delta {t^4}}}{4}{I_2}}&{\frac{{\delta {t^3}}}{2}{I_2}}\\
{\frac{{\delta {t^3}}}{2}{I_2}}&{\delta {t^2}{I_2}}
\end{array}} \right]\\
H =& \left[ {\begin{array}{*{20}{c}}
	{{I_2}}&{{0_2}}
	\end{array}} \right]\qquad~~~	R = \sigma _\varepsilon ^2{I_2}
\end{align*}
where ${I_2}$ represents the $2\times 2$ unit matrix, $0_2$ represents the $2\times 2$ zero matrix, $\sigma _v^2 = 1m{s^{ - 2}}$, $\sigma _\varepsilon ^2 = 2m{s^{ - 2}}$, and {\color{black}$\delta t = 1s$ denotes the sampling period}. The surviving probability is given as a constant ${p_{S}} =0.99$. The detection probability is unknown and given as ${p_{D}} =0.98$. {\color{black} The number of clutter per scan is Poisson distributed of $\lambda_c =20$, uniformly distributed in region $S=\left[ {{\rm{ - 2000}},{\rm{2000}}} \right]m\times\left[ {{\rm{ 0}},{\rm{2000}}} \right]m$. }The initial models for ten targets are given in Table \ref{Target States} and the death time here refers to the last time a target exists.
\par Besides, the expansion coefficient of the Beta distribution is given as $|k_\beta|=1.05$. The birth process is Poisson with parameters ${J_\gamma }=4$, ${\omega_\gamma }=0.01$, ${\widehat P_\gamma }=\text{diag}([50,50,50,50]^2)$. For each $j \in \left\{ {1,2,3,4} \right\}, \widehat m_{\gamma,1}^k=[-1500,250,0,0]^\top,  \widehat m_{\gamma,2}^k=[-250,1000,0,0]^\top,  \widehat m_{\gamma,3}^k=[250,750,0,0]^\top,  \widehat m_{\gamma,4}^k=[1000,1500,0,0]^\top$. In general cases, we are more interested in targets with detection probabilities greater than 0.5. Therefore, the Beta distribution factors of birth trajectory are given as $u=8$ and $v=2$, which means the initial value of detection probability is 0.8. The value of the $L$-scan approximation is set as $L=5$.
\begin{table}[!t]
	\centering
	\caption{The Initial Target States}
	\label{Target States}
	\scriptsize
	\begin{tabular}{c|c|c|c}
		\hline
	     \hline
		&Kinematic State&Birth Time$/s$&Death Time$/s$\\
		 \hline
		Target 1&$\left[1005,1489,8,-10\right]^{\top}$&1&100\\	 \hline
		Target 2&$\left[-256,1011,20,3\right]^{\top}$&10&100\\	 \hline
		Target 3&$\left[-1507,257,11,10\right]^{\top}$&10&100\\
		 \hline
		Target 4&$\left[-1500,250,43,0\right]^{\top}$&10&66\\	 \hline
		Target 5&$\left[246,735,15,5\right]^{\top}$&20&80\\  \hline
		Target 6&$\left[-243,993,-6,-12\right]^{\top}$&40&100\\
		 \hline
		Target 7&$\left[1000,1500,1,-10\right]^{\top}$&40&100\\	
		 \hline
		Target 8&$\left[250,750,-45,10\right]^{\top}$&40&80\\
		 \hline
		Target 9&$\left[1000,1500,-50,0\right]^{\top}$&60&100\\	
		 \hline
		Target 10&$\left[250,750,-40,25\right]^{\top}$&60&100\\
	 \hline
	\end{tabular}
\end{table}
\begin{figure}[!t]
	\centering
	\includegraphics[width=3.5in]{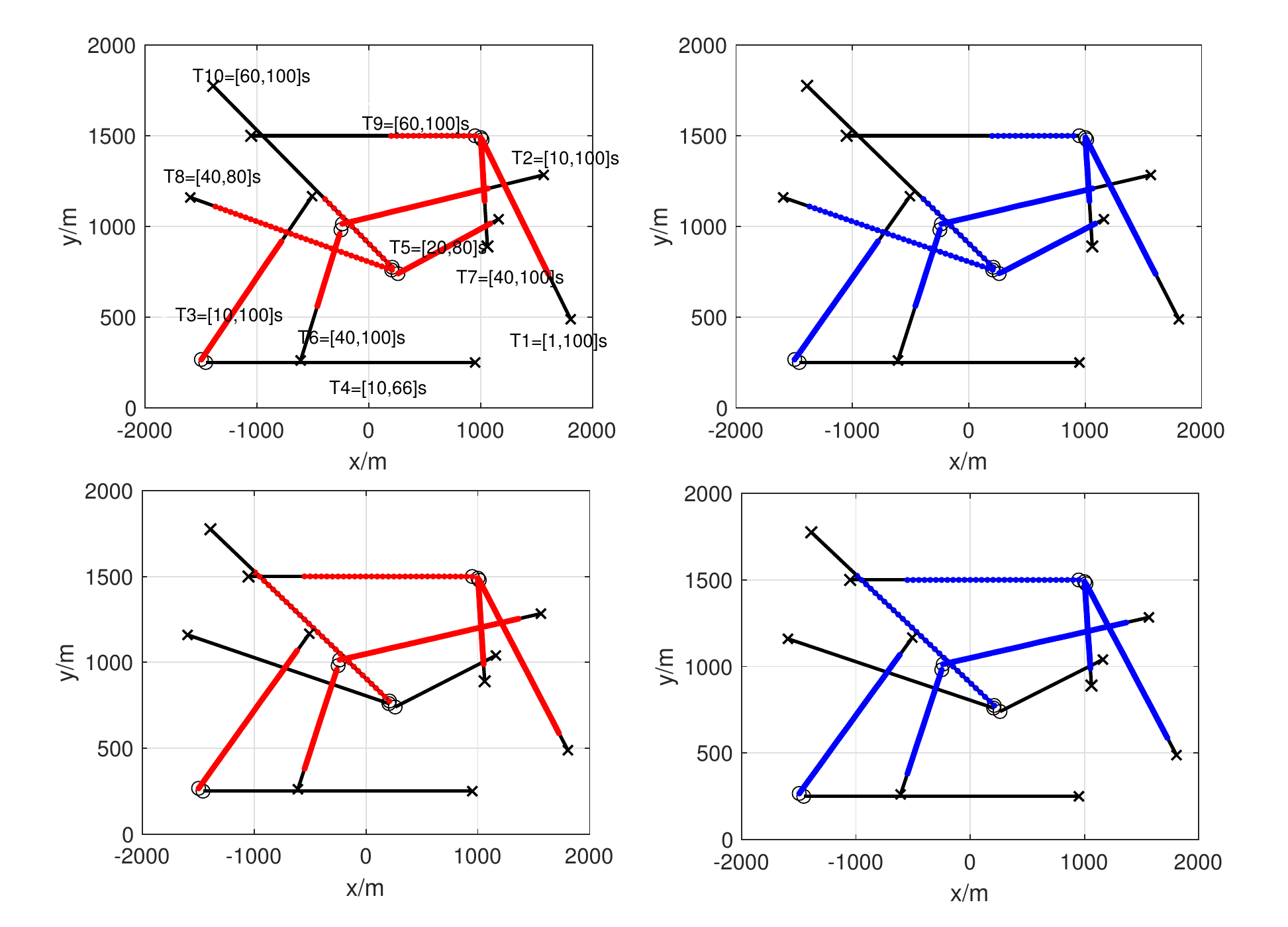}
	\caption{The trajectory of BG-U-TCPHD (left) at 75s (top), at 90s (bottom), and the trajectory of BG-U-TPHD (right) at 75s (top), at 90s (bottom) with Beta distribution factors $u=8, v=2$.}
	\label{11states}
\end{figure}
\par{\color{black} Besides, the weight threshold of pruning is given as
$\Gamma_p=10^{-5}$, the threshold of absorption is given as $\Gamma_a=4$ and the maximum of components is limited to $J_{max}=100$. For the BG-U-TCPHD filter, the cardinality distribution is capped at $N_{max}=100$.}
\par By running 1500 Monte Carlo realizations, the performance of the BG-U-TPHD and BG-U-TCPHD filters are obtained as follows. {\color{black}In order to do so, the error $d^2(\mathbf{X}_k^s,\mathbf{X}_k)$ at time $k$ between the estimated
alive set of trajectories $\mathbf{X}_k^s$ and the truth $\mathbf{X}_k$ are measured by the metric
for sets of trajectories with parameters $p = 2, c = 10, \gamma  = 1$, which is known as the trajectory metric (TM) error \cite{Angel2020TM}. It is also considered to normalize the error by the corresponding
time window $k$, thus, the root mean square (RMS) TM error $d(k)$ at
the time $k$ is obtained by
\begin{align}
d(k)=\sqrt{\frac{1}{N_{mc}}\sum_{i=1}^{N_{mc}}d^2(\mathbf{X}_{k,i}^s,\mathbf{X}_k)/k},
\end{align}
where $\mathbf{X}_{k,i}^s$ denotes the estimation of sets of alive trajectories at time $k$ in the $i-th$ Monte Carlo run. The TM error also includes the error for the localization of detected targets, false targets, missed targets and track switches.}
\begin{figure}[!t]
	\centering
	\includegraphics[width=3.5in]{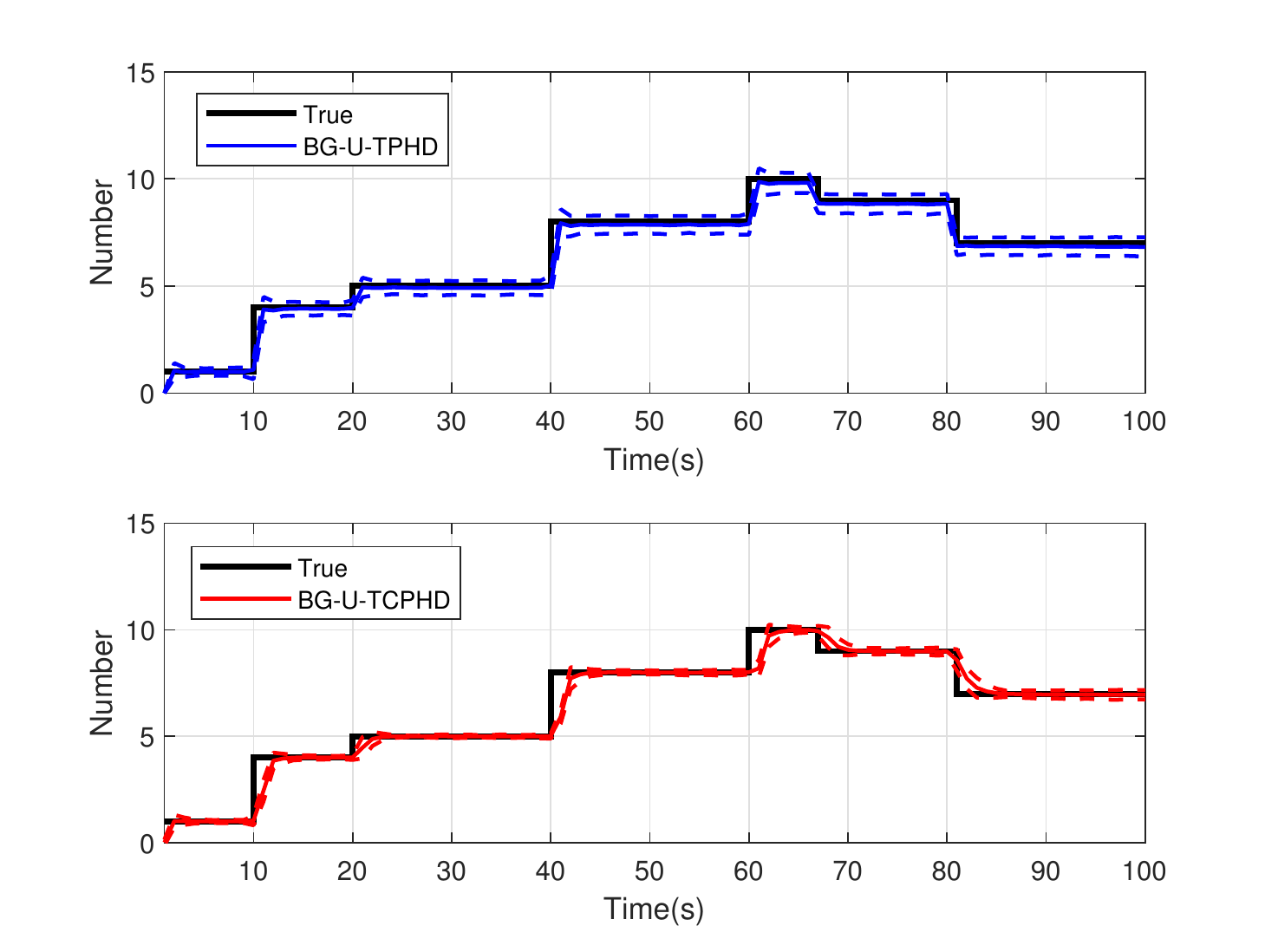}
	\caption{The BG-U-TPHD filter (top) and the BG-U-TCPHD filter (bottom) with Beta distribution factors $u=8,v=2$, the solid lines represent the number estimation, and the dashed lines represent estimation after calculating the standard deviation.}
	\label{11numbers}
\end{figure}
\begin{figure}[!t]
	\centering
	\includegraphics[width=3.5in]{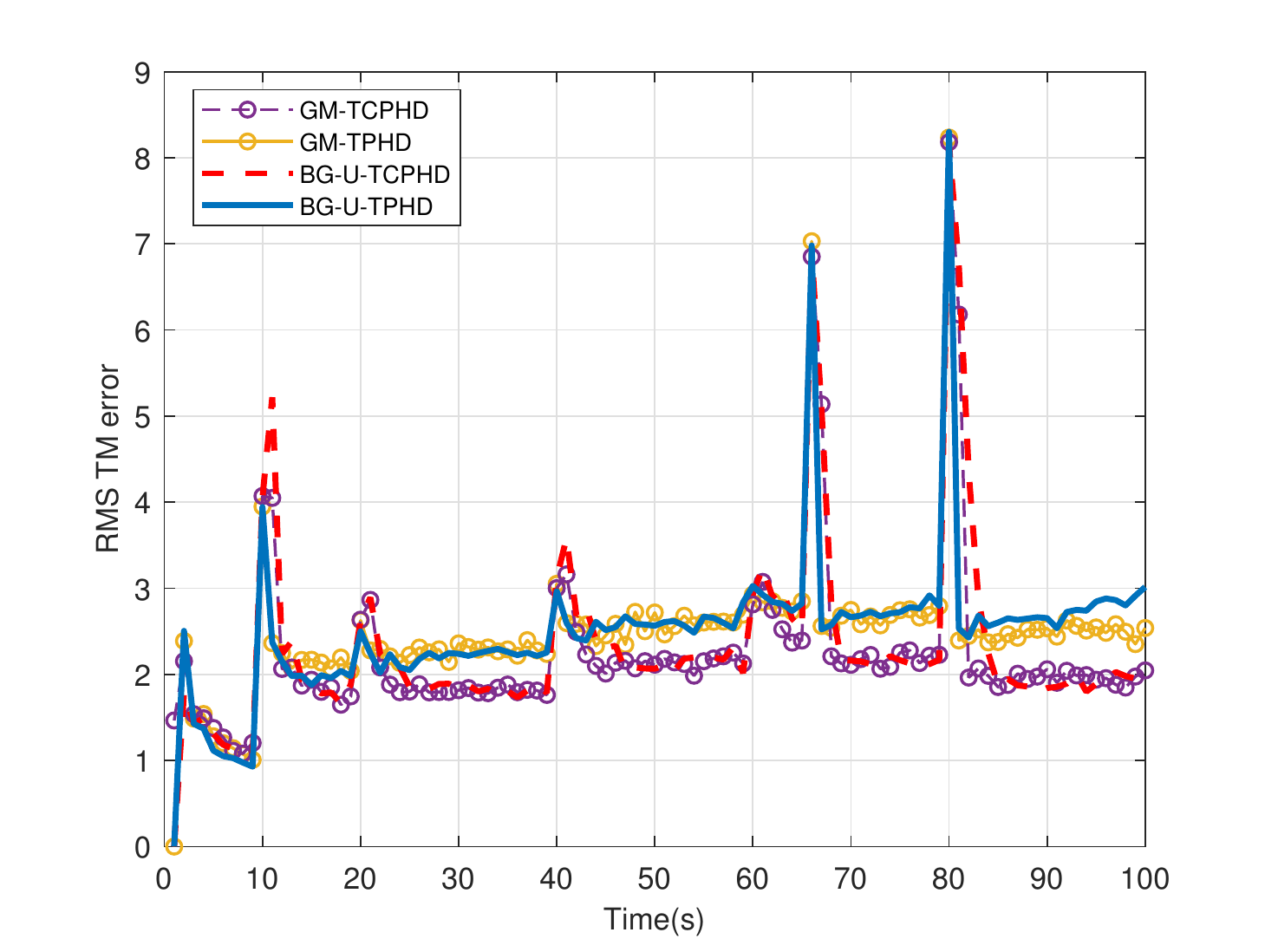}
	\caption{The RMS trajectory metric error for the BG-U-TPHD and BG-U-TCPHD filters with $u=8, v=2$, as well as the GM-TPHD and GM-TCPHD filters with known detection probability.}
	\label{11TM}
\end{figure}
\begin{figure}[!t]
	\centering
	\includegraphics[width=3.6in]{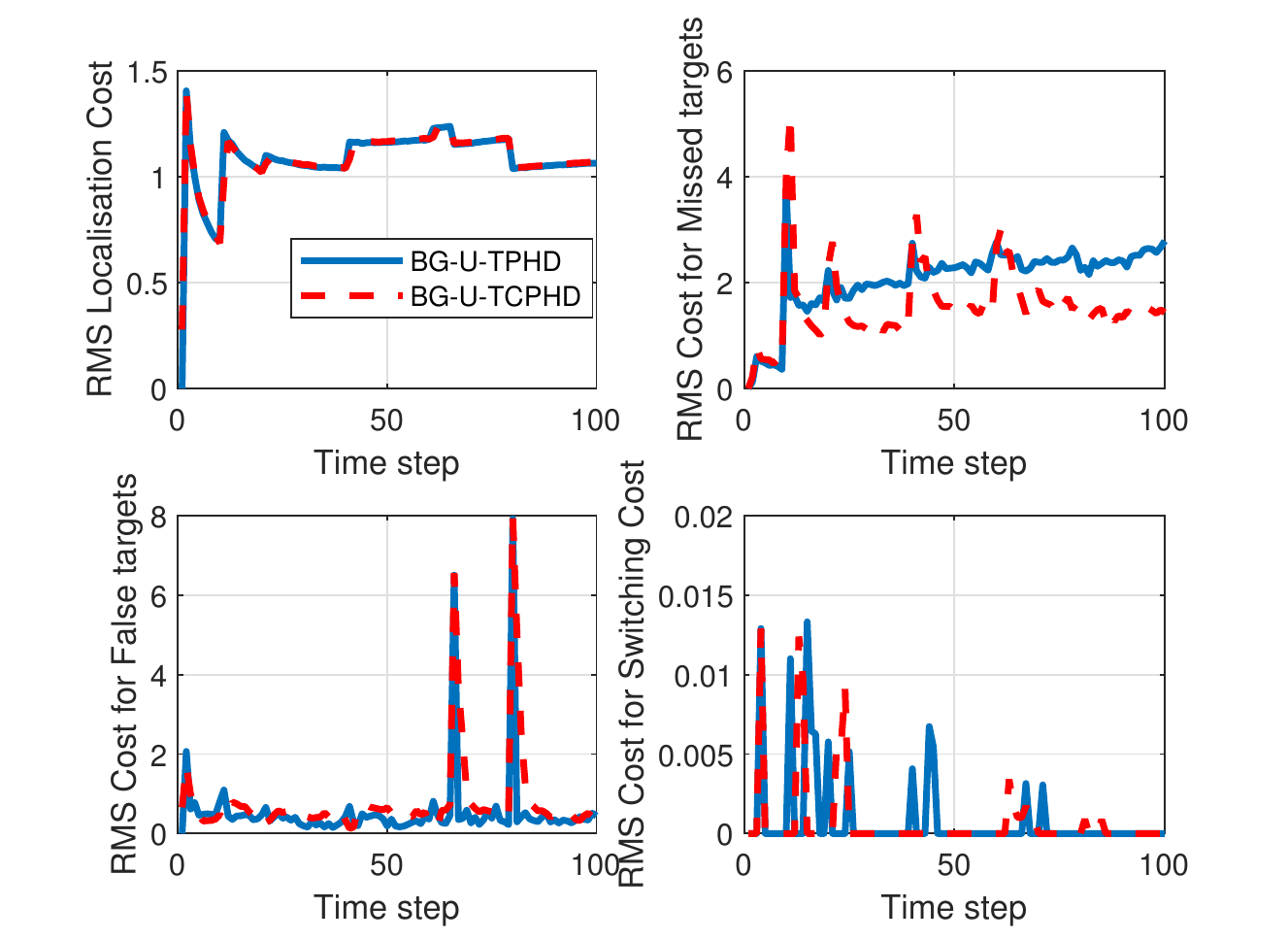}
	\caption{{\color{black}The decomposition of the RMS trajectory metric error for the BG-U-TPHD and BG-U-TCPHD filters with $u=8, v=2$, including the error for the localization of detected targets, false targets, missed targets and track switches.}}
	\label{TM_fenjie}
\end{figure}
\par It can be seen from Figs. \ref{11states}--\ref{11TM} that, both the BG-U-TPHD and BG-U-TCPHD filters can obtain excellent trajectory estimation which is {\color{black}close to the GM-TPHD and GM-TCPHD filters with the known detection profile \cite{Angel2019TPHD}, while the BG-U-TCPHD filter performs} much better than the BG-U-TPHD filter. {\color{black}The reason is that the BG-U-TCPHD filter also propagates the cardinality distribution but the BG-U-TPHD filter does not.} The decomposition of the RMS trajectory metric error for proposed filters is shown in Fig. \ref{TM_fenjie}. It shows that both filters possess similar error for the localization and false targets, and they nearly have no track switch error. However, the error for missed targets in the BG-U-TCPHD filters is much lower than that in the BG-U-TPHD filter.


\par
\begin{figure}[!t]
	\centering
	\begin{minipage}[t]{0.48\textwidth}
	\centering
	\includegraphics[width=3.5in]{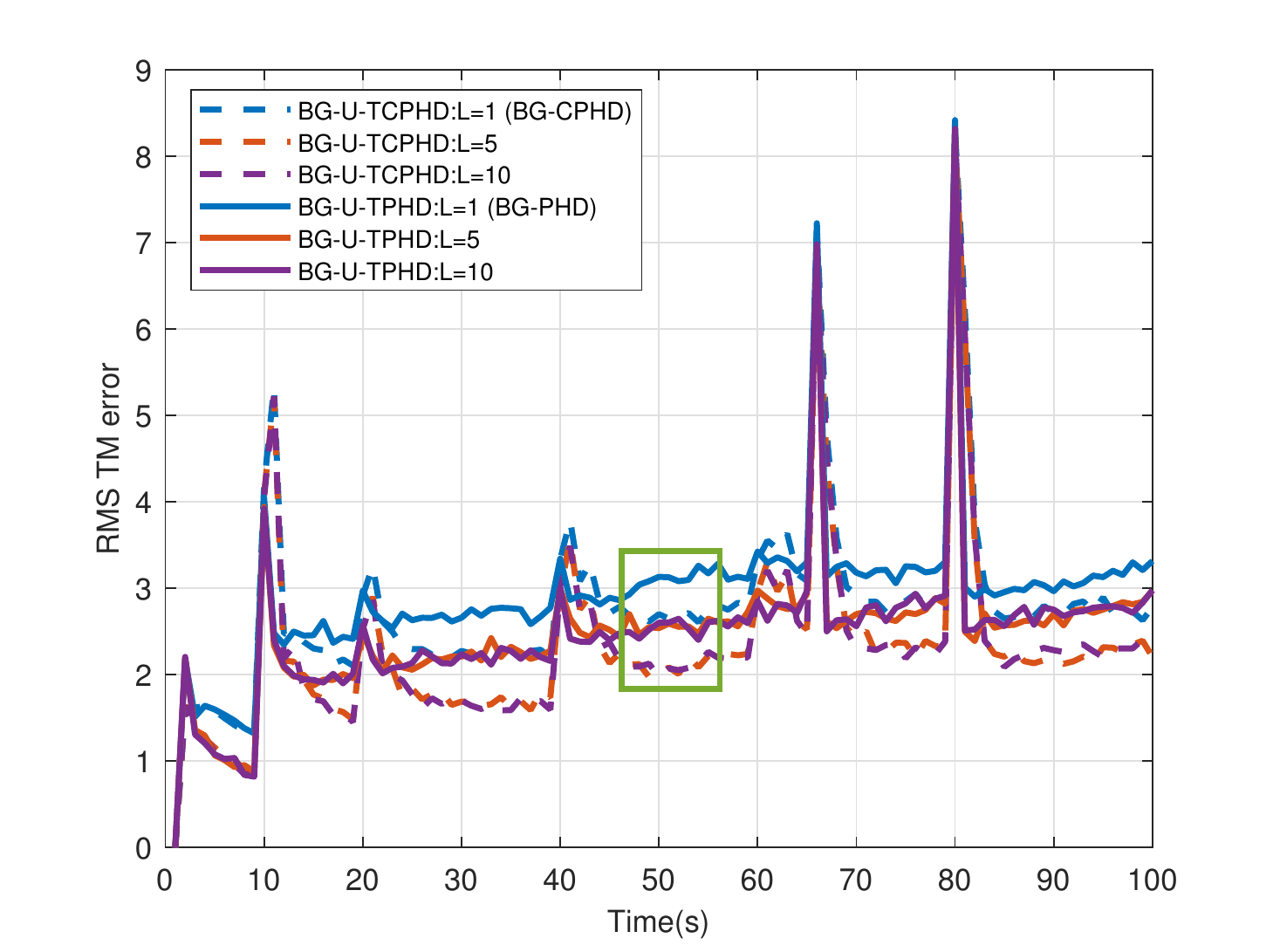}
     \label{fig:side:a}
	\end{minipage}
	\begin{minipage}[t]{0.48\textwidth}
    \centering
	\includegraphics[width=2.8in]{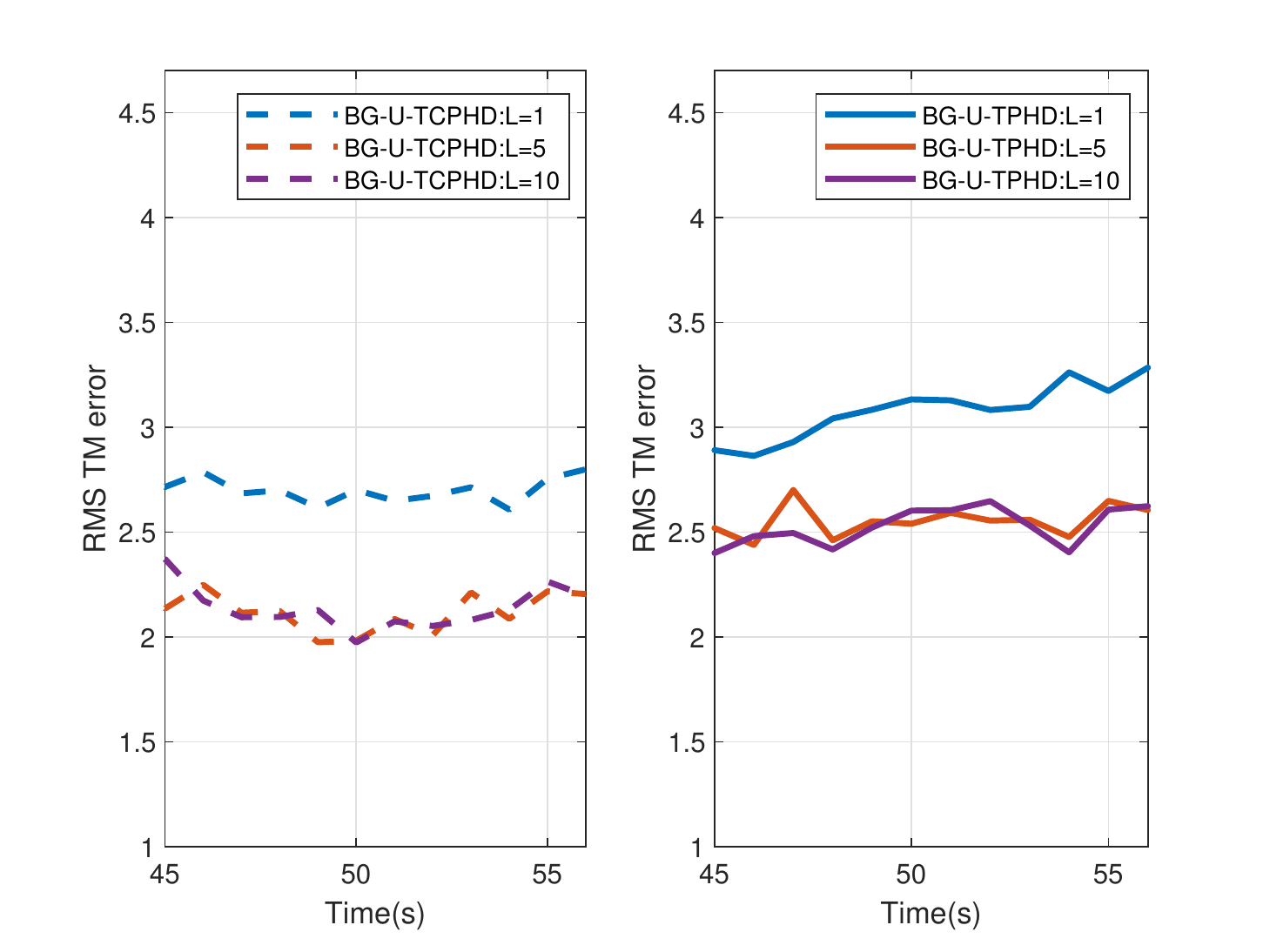}
   \label{fig:side:b}
    \end{minipage}
	\caption{{\color{black}The RMS TM error for the BG-U-TPHD and BG-U-TPHD filters with Beta distribution factors $u=8, v=2$ under the models of $L = 1, 5, 10$. The partial enlarged figure of the BG-U-TCPHD filter is shown in the lower left and the BG-U-TPHD filter is shown in the lower right}}
	\label{LscanTM}
\end{figure}

 The value of the $L$-scan approximation is of great importance in both filters, so the influence of different values of the $L$-scan will be compared in detail. It can be seen from Fig. \ref{LscanTM} that, with different values of the $L$-scan approximation, the BG-U-TCPHD filter always performs better than the BG-U-TPHD filter. Besides, it should be noted that, for both filters, the RMS TM error decreases when the $L$ increases, and the amount reduced gradually becomes less and less. When the $L$ is bigger than 5, the performance nearly has no improvement. Meanwhile, the computational burden due to the increasing value of the $L$-scan approximation should also be concerned. The averaged times to run one Monte Carlo iteration with a 2.8 GHz Intel i7 laptop are listed in Table \ref{Lscan}.
\begin{table}[!t]
	\centering
	\caption{The Run Time$(s)$ of The BG-U-TPHD and BG-U-TPHD Filters}
	\label{Lscan}
	\footnotesize
	\begin{tabular}{cccccccc}
		\hline
		\hline
		$L$&1&2&5&10&15&30&60\\
		\hline
		BG-U-TPHD&3.56&3.60&3.75&4.75&7.81&14.9&66.8\\
		BG-U-TCPHD&4.12&4.13&4.25&5.15&8.12&15.5&68.1\\
			\hline
	\end{tabular}
\end{table}
\par From Table \ref{Lscan}, when the $L$ is less than 5, the decrease of the $L$ cannot effectively improve the calculation speed, while the computational burden sharply increases if the $L$ is bigger than 10. In terms of both computational burden and accuracy, the $L$ is usually a much smaller quantity than the length of the trajectory. Therefore, in the Scenario 1, the $L=5$ is a suitable choice for both filters. {\color{black}Please note that, in Fig. \ref{11TM}, when $L=1$, the BG-U-TPHD and BG-U-TCPHD filters are equivalent to the BG-PHD and BG-CPHD filters \cite{Mahler2011UPHD} in performance.} 
\subsection{Scenario 2}
\begin{figure}[!t]
	\centering
	\includegraphics[width=3.5in]{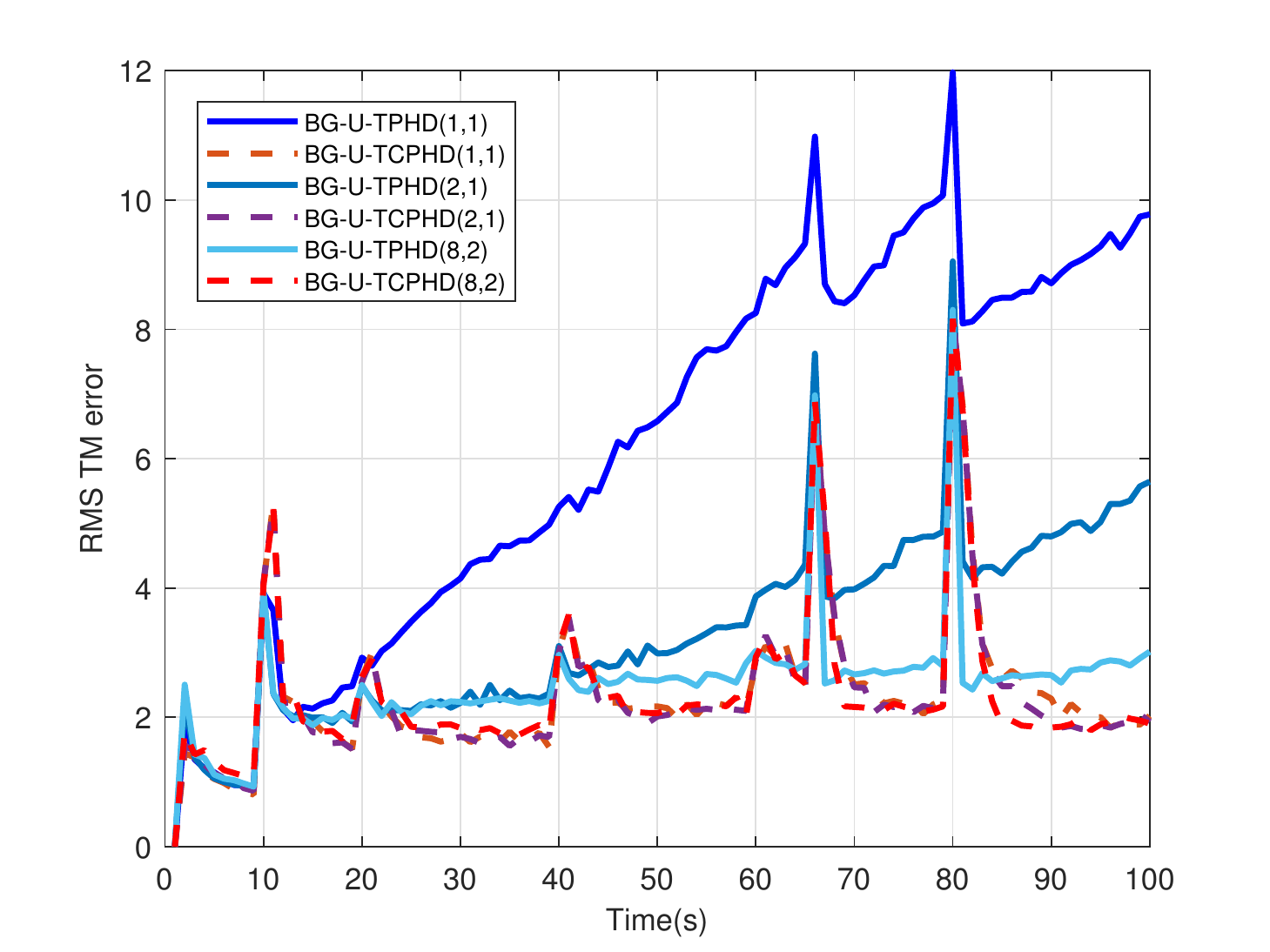}
	\caption{The RMS trajectory metric error for the BG-U-TPHD and BG-U-TPHD filters with different Beta distribution factors ($L$=5).}
	\label{CdB}
\end{figure}

In this section, we will focus on the performance of the BG-U-TPHD and BG-U-TCPHD filters with different Beta distribution factors, lower detection probability and uneven detection profile. First, the influence of different Beta distribution factors is elaborated and the uniform detection profile is adopted as $p_D=0.98$ being the same as the Scenario 1. 
When $u=1$ and $v=1$, the Beta distribution is equal to the uniform distribution. It can been seen from Fig. \ref{CdB} that, both filters can respond faster to the change of cardinality and produce a smaller error when the initialization for detection probability {\color{black} is closer} to the truth, while the BG-U-TPHD filter performs divergence when the initialization is far from the truth. In contrast, the BG-U-TCPHD filter is more robust.
\par Note that, compared to the robust BG-CPHD and BG-PHD filters in \cite{Mahler2011UPHD}, the initialization of the Beta distribution factors is more important for the BG-U-TPHD and BG-U-TCPHD filters, especially for the former. Because the trajectory state of Beta-Gaussian component with the smaller weight is directly abandoned, as indicated by Table \ref{Algorithm}. The trajectory estimation accuracy will degrade in the case of inaccurate estimation of detection probability. While the merging procedure in the robust BG-CPHD and BG-PHD filters \cite{Mahler2011UPHD} can reduce this influence to some extent.
\begin{table}[!t]
	{
		\begin{center}
			\caption{Average TM errors for the BG-U-TPHD and BG-U-TCPHD filters with different detection probabilities}
			\footnotesize
			\label{lowPD}
			\begin{tabular}{c|ccc|ccc}
				\hline
				\hline
				&\multicolumn{3}{c|}{ BG-U-TPHD}&\multicolumn{3}{c}{ BG-U-TCPHD}\cr
				\hline
				($u$,$v$)&(1,1)&(2,1)&(8,2)&(1,1)&(2,1)&(8,2)\cr
				$p_D=0.98$&6.41&3.23&2.45&2.20&2.15&2.08\cr

				$p_D=0.85$&11.51&9.80&6.15&4.24&4.04&3.80\cr

				$p_D=0.73$&16.53&13.17&13.13&5.20&5.06&5.04\cr
					\hline
			\end{tabular}
	\end{center}}
\end{table}
\par For targets with lower detection probabilities, from Table \ref{lowPD}, the decrease of detection probability seriously worsens the performance of the BG-U-TPHD filter, but has a smaller effect on the BG-U-TCPHD filter. 
\begin{table}[!t]
	\centering
	\caption{The Initial Target States}
	\label{BG-U-TCPHD}
	\scriptsize
	\begin{tabular}{c|c|c|c}
		\hline
		\hline
		&State&Birth Time$/s$&Death Time$/s$\\
		\hline
		Target 1&$[1005,1489,8,-10]^{\top}$&1&100\\	 \hline
		Target 2&$\left[-256,1011,20,3\right]^{\top}$&20&80\\	 \hline
		Target 3&$\left[-1507,257,11,10\right]^{\top}$&30&100\\
		\hline
	\end{tabular}
\end{table}
\begin{figure}[!t]
	\centering
	\includegraphics[width=3in]{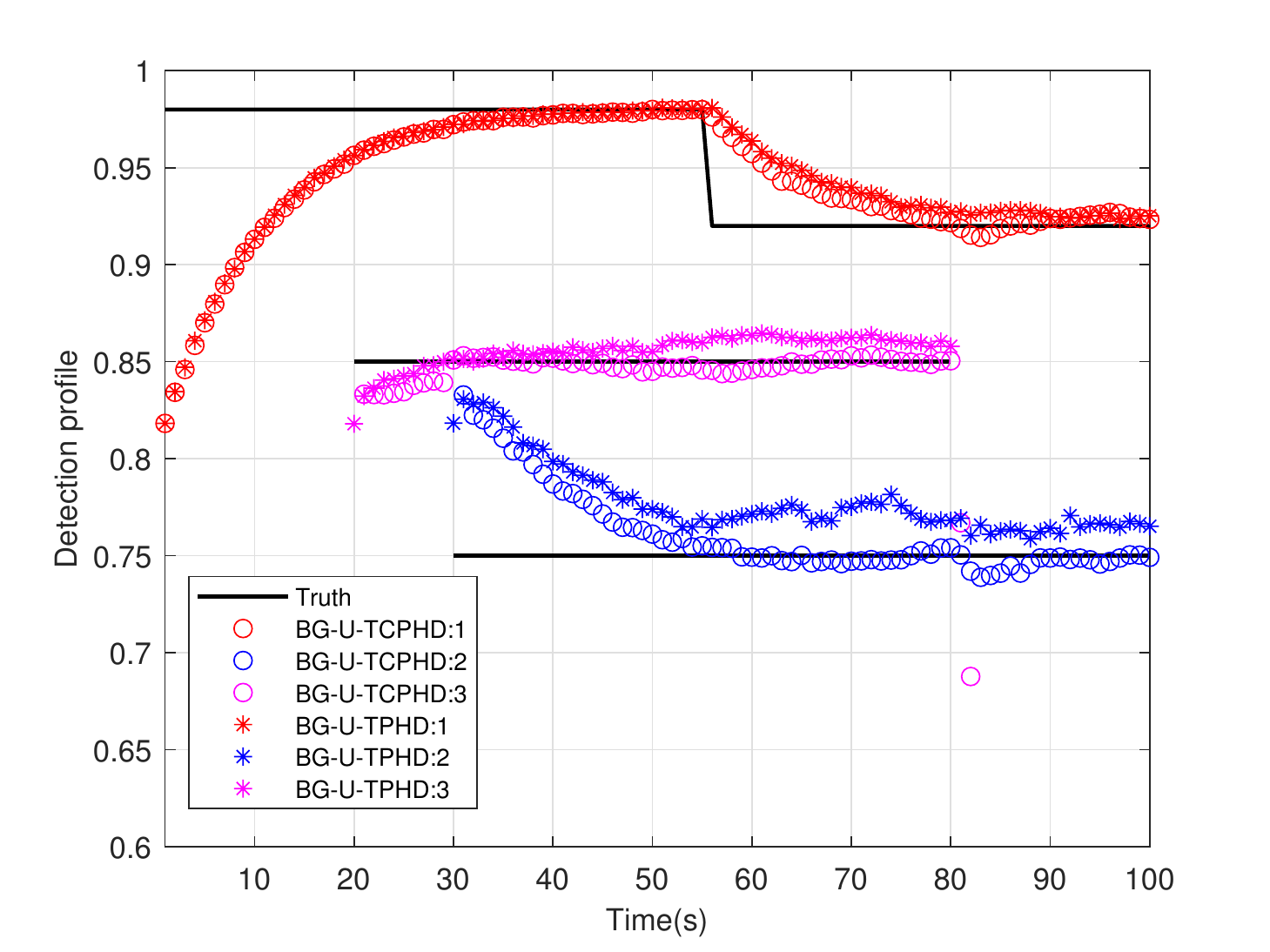}
	\caption{The average estimation of the uneven detection profile for the BG-U-TPHD and BG-U-TCPHD filters. The detection profile of different targets are distinguished with different color. The purple circles over $80$s are caused by the hysteresis of the BG-U-TCPHD filter for death trajectories.}
	\label{PD}
\end{figure}
For the uneven detection profile case, the Scenario 1 is simplified to three targets with different detection probabilities, which are given by the equation 
\begin{equation}\label{equ_unpd}
		p_{D,j,k} =	\left\{
		\begin{array}{lr}
			0.98, &j=1,k\le55 \\
			0.92, &j=1,k>55 \\
			0.85, &j=2\\
			0.75, &j=3
		\end{array}
		\right.
\end{equation}
where $j\in\{1,2,3\}$ denotes different targets and $k$ denotes the time. The initial target states are listed in Table \ref{BG-U-TCPHD} and rest parameters are the same as Scenario 1. It can be seen from Fig. \ref{PD} that there is an initial settling in period, but after this miss distance, the estimation of detection profile converges to the truth and performs fluctuation with the changes of the number of targets. It turns out that the proposed BG-U-TPHD and BG-U-TCPHD filters provide the satisfactory performance at adaptively learning uneven detection profile, while the BG-U-TCPHD filter performs better.

\section{Conclusion}
 In this paper, we derived the recursive equations of the U-TPHD and U-TCPHD filters by using the KLD minimization. Both filters can perform robustly in scenarios with unknown and time-varying target detection probability. Meanwhile, for the computation efficiency, the analytic recursions are also presented of the U-TPHD and U-TCPHD filters, which consider the effect of the unknown detection probability for only the latest frame time. The Beta-Gaussian mixture method is also adopted for the implementation of the proposed filters, which are referred to as the BG-U-TPHD and BG-U-TCPHD filters. Besides, the $L$-scan approximations of these filters with much lower computational burden are also presented. Finally, simulation results demonstrate that the BG-U-TPHD and BG-U-TCPHD filters can achieve robust tracking performance to adapt to unknown detection profile. Besides, it also shows that usually a small value of the $L$-scan approximation can achieve almost full efficiency of both filters but with a much lower computational costs.

\appendices
\section{Proof of Prediction}{\color{black}
This section aims to clarify the relationship between the transition density of the augmented trajectory $\tilde X=(t,x^{1:i},a^{1:i})$ and the transition density of targets $x$ and detection probability $a$ of a single frame time. Taking the U-TPHD filter for example, the augmented trajectory state at time $k-1$ is given as $\tilde{\underline{X}}=(\underline{t},\underline{x}^{1:\underline{i}},\underline{a}^{1:\underline{i}})$ and the predicted augmented trajectory state at time $k$ is given as  $\tilde{X}=(t,{x}^{1:i},{a}^{1:i})$. Thus, the prediction of the U-TPHD filter is obtained by
\begin{align}
		&{D_{k|k - 1}}\left( {\tilde X} \right)\\
		 =&{\gamma}\left( {\tilde X} \right)\delta_{1}[i]\delta_{k}[t] + \int{p_{S,k}}\left( \tilde{\underline{X}} \right)\tilde f\left( \tilde{X}|\tilde{\underline{X}} \right){D_{k - 1}}\left( \tilde{\underline{X}} \right)d\tilde{\underline{X}}\notag,
\end{align}
with only considering the first-order Markov process and alive trajectories:
\begin{align}
	&{D_{k|k - 1}}\left( {\tilde X} \right)\\
	=&{\gamma}\left(t,x^{1:i},a^{1:i} \right)\delta_{1}[i]\delta_{k}[t] + {p_{S,k}}\left( x^{i-1} \right)f\left( x^i,a^i|x^{i-1},a^{i-1} \right)\notag\\
	&\times\iint\delta_{\underline{x}^{1:\underline{i}}}({x}^{1:i-1})\delta_{\underline{a}^{1:\underline{i}}}({a}^{1:i-1})\delta_{\underline{t}}[{t}]\delta_{\underline{i}+1}[{i}],\notag\\
	&\times {D_{k - 1}}\left(\underline{t},\underline{x}^{1:\underline{i}},\underline{a}^{1:\underline{i}}\right)d\underline x^{1:\underline{i}}d\underline a^{1:\underline{i}}\notag\\
	=&{\gamma _k}\left( {\tilde X}\right) + {p_{S,k}}\left( x^{i-1} \right)f\left( x^i,a^i|x^{i-1},a^{i-1} \right)\notag\\
	&\times{D_{k - 1}}\left(t,{x}^{1:i-1},{a}^{1:i-1}\right),\notag
\end{align}
where $\tilde f( \tilde{X}|\tilde{\underline{X}})$ and $f( x^i,a^i|x^{i-1},a^{i-1})$ denote the transition density of augmented trajectory and that of augmented targets, respectively, and $t \in \{1,2,...,k-1\}$ and $t+i-1=k$. As indicated by the equation \eqref{trans}, we consider the switch of detection probability is independent of the trajectory state, thus, the predicted PHD $D_{k|k-1}(\tilde X)$ can be simplified as, which is the same as equation \eqref{equ_UTPHD_pr_total}
\begin{align}
		D_{k|k-1}(\tilde X)=&{\gamma _k}( \tilde X ) +{p_{S,k}}\left( x^{i-1} \right)f\left( x^i|x^{i-1},a^{i-1} \right)\notag\\
	&\times g\left( a^i|a^{i-1} \right){D_{k - 1}}\left(t,{x}^{1:i-1},{a}^{1:i-1}\right),\notag
\end{align}
where $g (\cdot|\cdot)$ denotes the transition density of detection probability.
The prediction of the U-TCPHD filter enjoys the same principle but also considers the cardinality distribution, which is a simple extension of \cite{Angel2019TPHD}. So it is omitted here.}
\section{Proof of Proposition 2}
In this appendix, the proof of Proposition \ref{UTPHD_up} is elaborated. The proof of KLD minimization is omitted, because the trajectory augmented with a detection probability sequence is a simple extension of the classical trajectory state, and the proof of the latter can be found in \cite{Angel2019TPHD}. Thus, the aim is to compute the updated PHD of augmented trajectory density by Bayes rule. First, the density of measurements of augmented targets at time $k$ is given as
\begin{align}
\ell^\tau_k&(Z_k|\{\tilde{x}_1,...,\tilde{x}_n\} ) \notag\\
=& {e^{ - {\lambda _c}}}\left[ {\prod\limits_{j = 1} {{\lambda _c}\bar c } ({z_j})} \right]\left[ {\prod\limits_{j = 1}^n {(1 - a_j)} } \right]\notag\\
&\times\sum\limits_{\sigma  \in {\Gamma _{n}}} {\prod\limits_{j:\sigma_j>0} {\frac{{a_j\cdot l_k({z_{\sigma_j}}|{x_j})}}{{(1 -a_j){\lambda _c}\bar c ({z_{{\sigma _j}}})}}} },
\end{align}
where ${\Gamma _{n}}$ denotes all kinds of possible associations between $n$ targets and a set of measurements $\sigma$. The notation $\sigma_j> 0$ indicates the target $j$ is detected and associated with measurements $\sigma_j$. 
$a$ denotes $a^{k-t+1}$, which represents the detection profile at time $k$. Given the augmented trajectory RFS $\mathbf{\tilde{X}}_k=\{\tilde{X}_1,...,\tilde{X}_n\}$ at time $k$,
the posterior PHD of the density of the augmented trajectories can be obtained by Bayes rule
\begin{align}
D_{k}&(\tilde{X})\notag\\
=&\frac{1}{{{\ell_k}\left( {{Z_k}} \right)}}\int{{\ell_k}\left( {{Z_k}|\{\tilde{X}\}\cup\mathbf{\tilde{X}}_k} \right){p_{k|k - 1}}\left( \{\tilde{X}\}\cup\mathbf{\tilde{X}}_k \right)}\delta\mathbf{\tilde{X}}_k\notag\\
=&\frac{1}{{{\ell_k}\left( {{Z_k}} \right)}}\sum_{n=0}^{\infty}\frac{1}{n!}\int{{\ell_k}\left( {{Z_k}|\{\tilde{X},\tilde{X}_1,...,\tilde{X}_n\}} \right)}\notag
\end{align}
\begin{align}
&{\times{p_{k|k - 1}}\left( \{\tilde{X},\tilde{X}_1,...,\tilde{X}_n\} \right)}\delta\tilde{X}_{1:n}\notag\\
=&\frac{D_{k|k-1}(\tilde X)}{{{\ell_k}\left( {{Z_k}} \right)}}\sum_{n=0}^{\infty}\frac{1}{n!}\int{{\ell^\tau_k}\left( {{Z_k}|\{\tilde{x},\tilde{x}_1,...,\tilde{x}_n\}} \right)}\notag\\
&{\times e^{-\lambda}\prod_{j=1}^{n}D_{k|k-1}^{\tau}(\tilde x_j)}\delta\tilde{x}_{1:n},
\end{align}
where
\begin{align*}
D_{k}^\tau (x,a) =&\sum\limits_{t = 1}^k {\iint {{D_{k}(t,{x^{1:k - t}},a^{1:k - t},x,a)d{a^{1:k - t}}d{x^{1:k - t}}} }},
\end{align*}
which is the PHD of the posterior density of augmented targets at time $k$ and integrating the {\color{black}sequence of both detection probabilities and target states.} The measurement set $Z_k$ is considered as a single frame time and comes from the clutter and targets. Therefore the denominator of Bayes rule is obtained as \cite{Angel2015KLD}.
\begin{align}
{{{\ell_ k}\left( {{Z_k}} \right)}}=&e^{-\lambda_c-\iint{a\cdot D^{\tau}_{k|k-1}(x,a)dadx}}\\
&\times\prod_{z\in Z_k}\left[\lambda_c\bar c(z)+\iint {a\cdot l(z|x)D^{\tau}_{k|k-1}(x,a)dadx}\right].\notag
\end{align}
\par Based on \cite{Angel2015KLD}, at time $k$, the target $x$ can be detected, with probability $a$ and miss detection with probability $1-a$. Therefore, we can obtain
\begin{align}
{\ell^\tau_k}&\left( {{Z_k}|\{\tilde{x},\tilde{x}_1,...,\tilde{x}_n\}}\right)\notag\\
=&(1-a)\cdot{\ell_k^\tau}\left( {Z_k}|\{\tilde x_1,...,\tilde x_n\}\right)\notag\\
&+a\sum_{z\in Z_k}{l_k}\left( {z}|x\right){\ell_k^\tau}\left( {Z_k \backslash \{z\}}|\{\tilde x_1,...,\tilde x_n\}\right).
\end{align}
\par The posterior PHD of the density of the augmented trajectories is
\begin{align}
{D_k}\left( {\tilde X} \right) =& {D_{k|k - 1}}\left( {\tilde X} \right)(1-a)\\
& + {D_{k|k - 1}}\left( {\tilde X} \right){a}\notag\\
&\times\sum\limits_{z \in {Z_k}} {\frac{{{l_k}(z|x)}}{{{\lambda _c}\bar c (z) + \iint a\cdot l_k(z|x)D^\tau_{k|k-1}(x,a)dadx }}}.\notag
\end{align}

\section{Proof of Proposition 4}
In this appendix, the proof of Proposition 4 is described in detail. Similar to Appendix A, the posterior PHD can be obtained by the Bayes rule
\begin{align}
D_{k}&(\tilde{X})\notag\\
=&\frac{1}{{{\ell_k}\left( {{Z_k}} \right)}}\sum_{n=0}^{\infty}\frac{1}{n!}\int{{\ell_k}\left( {{Z_k}|\{\tilde{X},\tilde{X}_1,...,\tilde{X}_n\}} \right)}\notag\\
&{\times{p_{k|k - 1}}\left( \{\tilde{X},\tilde{X}_1,...,\tilde{X}_n\} \right)}\delta\tilde{X}_{1:n}\notag\\
=&\frac{1}{{{\ell_k}\left( {{Z_k}} \right)}}\sum_{n=0}^{\infty}(n+1)\int{{\ell_k}\left( {{Z_k}|\{\tilde{X},\tilde{X}_1,...,\tilde{X}_n\}} \right)}\notag\\
&\times \rho_{k|k-1}(n+1)\bar p_{k|k-1}(\tilde X)\prod_{j=1}^{n}\bar{p}_{k|k-1}(\tilde{X}_j)\delta\tilde{X}_{1:n}.
\end{align}
where
\begin{align}
\int&{{\ell_k}\left( {{Z_k}|\{\tilde{X}_1,...,\tilde{X}_n\}} \right)\prod_{i=1}^{n}\bar{p}_{k|k-1}(\tilde{X}_i)\delta\tilde{X}_{1:n}}\notag\\
=& \prod\limits^{|Z_k|=M}_{j=1} {\bar c } (z_j)\sum\limits_{i = 0}^{\min (M,n)} {(M- i)!{\rho _{c,k}}(M- i)\frac{n!}{(n-i)!}}\notag\\
&\times \frac{\left[ {\iint {(1 - a)D_{k|k-1}^\tau(x,a) dadx}} \right]^{n - i}}{\left[\iint{D_{k|k-1}^\tau(x,a)dadx}\right]^n}\notag\\
&\times \sum\limits_{S_k \subseteq Z_k,|S_k| = i}
{\frac{\prod\limits^{i}_{j=1}\left[\iint{a\cdot l(z_j|x)D_{k|k-1}^\tau(x,a))dadx}\right]}
{\prod\limits^{i}_{j=1}\bar c (z_j)}}\notag\\
=&\prod_{z\in Z_k}{\bar c(z)\Upsilon_k^0\left[D_{k|k-1}^\tau ;Z_k\right](n)}.
\end{align}
\par Based on the theory of \cite{Angel2015KLD}, we can obtain
\begin{align}
&\sum_{n=0}^{\infty}{(n+1)\rho_{k|k-1}(n+1)}\int{\ell_k}\left( {{Z_k}|\{\tilde{X}_1,...,\tilde{X}_n\}} \right)\\
&\times\prod_{i=1}^{n}\bar{p}_{k|k-1}(\tilde{X}_i)\delta\tilde{X}_{1:n}\notag\\
&=\iint{D_{k|k-1}^\tau(x,a)dxda}\langle{\rho_{k|k-1},\Upsilon_k^1[D_{k|k-1}^\tau ;Z_k]\rangle}\prod_{z\in Z_k}{\bar c(z)}.\notag
\end{align}
\par The denominator of Bayes rule is
\begin{align}
{{{\ell_k}\left( {{Z_k}} \right)}}=\prod_{z\in Z_k}{\bar c(z)\langle{\rho_{k|k-1},\Upsilon_k^0[D_{k|k-1}^\tau ;Z_k]\rangle}}.
\end{align}
\par Therefore, the posterior PHD of the augmented multi-trajectory density is given as
\begin{align}
{D_k}\left( {\tilde X} \right) =& {D_{k|k - 1}}\left( {\tilde X} \right) ({1-a})\\
&\times\frac{{\left\langle {\Upsilon _k^1\left[ {D_{k|k-1}^\tau ;{Z_k}} \right],{\rho _{k|k - 1}}} \right\rangle }}{{\left\langle {\Upsilon _k^0\left[ {D_{k|k-1}^\tau ;{Z_k}} \right],{\rho _{k|k - 1}}} \right\rangle}}\notag\\
&+ {D_{k|k - 1}}\left( {\tilde X} \right){a}\notag\\
&\times\sum\limits_{z \in {Z_k}} {\frac{{l_k\left( {z|x} \right)}}{{\bar c\left( z \right)}}\frac{{\left\langle {\Upsilon _k^1\left[ {D_{k|k-1}^\tau ;{Z_k}\backslash \left\{ z \right\}} \right],{\rho _{k|k - 1}}} \right\rangle }}{{\left\langle {\Upsilon _k^0\left[ {D_{k|k-1}^\tau ;{Z_k}} \right],{\rho _{k|k - 1}}} \right\rangle }}}.\notag
\end{align}
\par Besides, the posterior cardinality distribution is also given by Bayes rule
\begin{align}
{\rho_{k}}(n)=& \frac{1}{{\ell(Z_k)n!}}\int {\ell(Z_k|\{ \tilde{X_1},...,\tilde{X_n}\} )}\notag\\
&{\times p_{k|k-1}(} \{\tilde{X_1},...,\tilde{X_n}\} )\delta\tilde{X}_{1:n}\notag\\
=& \frac{{\Upsilon _k^0\left[ {D_{k|k-1}^\tau ;{Z_k}} \right](n){\rho _{k|k - 1}}\left( n \right)}}{{\left\langle {\Upsilon _k^0\left[ {D_{k|k-1}^\tau ;{Z_k}} \right],{\rho _{k|k - 1}}} \right\rangle }}.
\end{align}

\bibliographystyle{IEEEtran}
\bibliography{UN}

\end{document}